\documentclass[a4paper,11pt]{article}
\pdfoutput=1 
\usepackage{jcappub} 

\usepackage{graphicx}
\usepackage{hyperref}
\usepackage{caption}
\usepackage{subcaption}

\title{Scalar-Tensor gravity with system-dependent potential and its relation with  Renormalization Group extended General Relativity}

\author[a]{Davi C. Rodrigues,}
\author[b]{Bertrand Chauvineau,}
\author[a]{Oliver F. Piattella}
\affiliation[a]{Departamento de F\'isica, Universidade Federal do Esp\'irito Santo, Av. F.Ferrari, 514, 29075-910, Vit\'oria, Brazil}
\affiliation[b]{Laboratoire Lagrange (UMR 7293), Universit\`e de Nice-Sophia Antipolis, CNRS, Observatoire de la C\^ote d'Azur, BP 4229, 06304 Nice cedex 4, France}

\emailAdd{davi.rodrigues@cosmo-ufes.org}
\emailAdd{Bertrand.Chauvineau@oca.eu}
\emailAdd{oliver.piattella@pq.cnpq.br}

\abstract{We show that Renormalization Group extensions of the Einstein-Hilbert action for large scale physics  are not, in general, a particular case of standard Scalar-Tensor (ST) gravity. We present a new class of ST actions, in which the potential is not necessarily fixed at the action level, and show that this extended ST theory formally contains the  Renormalization Group case. We also propose here a Renormalization Group scale setting identification that is explicitly covariant and valid for arbitrary relativistic fluids.}

\keywords{gravity, modified gravity, quantum field theory on curved space}

\begin{document}
\maketitle

\flushbottom

\section{Introduction}

Scalar-Tensor (ST) theories of gravity are known since the sixties, when  Brans-Dicke theory was proposed \cite{Brans:1961sx}. They can be defined as gravitational models, given by an action principle, whose fundamental fields are a rank two tensor (the metric) together with one or more scalar fields\footnote{The latter case is occasionally called ``multiscalar-tensor gravity", here the term ST is used for any number of scalar fields.}, and are such that they interact non minimally with the metric. They constitute theories with well defined actions that may work as extensions or alternatives to General Relativity. They are an ample framework that in particular includes the Horndeski action \cite{Horndeski:1974wa}, which has been in the focus of diverse current theoretical works (e.g., \cite{Deffayet:2013lga, Amendola:2012ky, Kase:2014cwa}). Moreover, purely metric gravitational theories, whose the kinetic part is nonlinear on the Ricci scalar (e.g., $f(R)$ theories) or nonlinear on other metric scalars (e.g., $f(R, R_{\mu \nu}R^{\mu \nu}, \Box R...)$), can be written as an action that depends linearly on these metric scalars but coupled to additional scalar fields, and hence in ST form (for a review on $f(R)$ see Ref.\cite{Capozziello:2010zz}, and for the more general case see Refs. \cite{Rodrigues:2011zi, Baykal:2013gfa}). 

In spite of the wide scope of  ST theories, we present here an extension to the standard picture. Its fundamental fields are indeed a metric and a scalar field (or some scalar fields), but the potential needs not to be fixed at the action level, which implies, as it will be shown, that its form can in general depend on constants associated with matter properties and the boundary conditions. In the standard ST picture, the potential is defined as a function of the scalar field, and this function is fixed at the action level (hence this function is the same for any system, say at the solar system or at cosmological scales). In the proposed extended picture, the potential is defined as a scalar whose spacetime dependence is neither explicit nor it is obtained from the metric or matter fields. This is a more general definition than the previous one, this case does not fix the form of the potential at the action level, it only states on what the potential cannot depend on. Implicitly, there is a statement that the potential, if not a constant, does depend on the scalar field, but there is no statement, at the action level, on the form of such dependence. At the field equations level, one may solve the differential equations in order to derive the specific potential form for the given matter content and boundary conditions.

Since the two type of potentials described above lead to different field equations solutions, it is convenient to introduce a notation in order to differentiate both cases. Namely, for the standard potential case, we use $V(\phi)$, while for the extended case we use $V\{\phi\}$. The full details on the system-dependent potential case can be found in section \ref{sec:STcomparison}.

\bigskip

In this work, we show that the  Renormalization Group (RG) extensions to the Einstein-Hilbert action, considering large scale phenomena, naturally can be put in the extended ST picture described above. This is a relevant result since it both provides a motivation to the extended ST that is here proposed and a way to compare gravity related RG effects to the ST picture.

Quantum effects in gravity have been studied for a long time within diverse approaches. Since General Relativity is a perturbatively non-renormalizable theory, it is common to consider it an effective low energy theory, therefore, at sufficiently high energies (or small distances), it should yield wrong predictions, and hence it should be modified (e.g., Ref. \cite{Stelle:1976gc, Donoghue:1993eb, Biswas:2011ar, Modesto:2014eta}). 

A particular type of quantum correction, and its phenomenological consequences, has been attracting considerable interest currently, namely that of nontrivial Renormalization Group (RG) flows.  These corrections can be relevant both in  quantum gravity  or in  Quantum Field Theory (QFT) in curved spacetime (for reviews on diverse aspects see \cite{Berges:2000ew, Niedermaier:2006wt, Shapiro:2008sf, Rosten:2010vm}). The interest on nontrivial RG flows comes from two fronts: one from a high energy perspective, and the other from a low energy one. Considering the former, nonperturbative renormalizability (with unitarity) may be achieved from the Asymptotic Safety program (for reviews see \cite{Niedermaier:2006wt, Percacci:2007sz, Ashtekar:2014kba}). The low energy case modifications can be motivated from the former high energy case, but do not really depend on it, in particular do not depend on the existence of a non-Gaussian UV fixed point which is crucial for the Asymptotic Safety program. It was realised  that there is actually no reason to assume that the $\beta$-functions\footnote{The $\beta$-function of a coupling constant $K$ is defined by $\beta_K = \mu \, d K /d\mu$, where $\mu$ is the RG scale.} of both the gravitational coupling $G$ and the cosmological constant $\Lambda$ must quickly approach zero as the RG energy scale becomes small, likewise happens to QED or QCD (e.g., \cite{Gorbar:2002pw, Shapiro:2009dh}). Hence, in particular, the RG flows of $G$ and $\Lambda$ need not to satisfy the Appelquist-Carazzone decoupling theorem \cite{Appelquist:1974tg}. Related comments and results can be found in diverse references, including Refs. \cite{Antoniadis:1991fa, Tsamis:1994ca, Shapiro:1994sz, Shapiro:2004ch, Reuter:2007de, Manrique:2008zw}. This motivated the search for consistent, and phenomenological relevant, effective classical descriptions that take into account certain running of $G$ and $\Lambda$ at systems that are ``large'' and usually considered to be fully classical. Many examples can be cited for the latter category, e.g. \cite{Reuter:1986wm, Goldman:1992qs, Bertolami:1993mh, Bottino:1995sa, Reuter:1996cp, Bonanno:2001hi, Bonanno:2001xi, Bentivegna:2003rr, Reuter:2003ca, Reuter:2004nv, Shapiro:2004ch, Reuter:2004nx, Babic:2004ev, Bauer:2005rpa, Brownstein:2005zz, Grande:2007wj, Borges:2007bh, Rodrigues:2009vf, Grande:2010vg, Domazet:2010bk, Domazet:2012tw, Koch:2010nn, Cai:2011kd, Ahn:2011qt, Costa:2012xw, Hindmarsh:2012rc, Perico:2013mna, Rodrigues:2014xka, Lima:2014hia, Koch:2014joa}. Some of these references consider the RG effects at the action level, which is the main case dealt in this work\footnote{This approach is called ``RG improved action'' in Ref. \cite{Reuter:2003ca}, in contraposition to other possibilities which are named ``RG improved equations'' and ``RG improved solutions''.} \cite{Reuter:2003ca, Reuter:2004nv, Shapiro:2004ch, Reuter:2004nx, Rodrigues:2009vf, Koch:2010nn, Cai:2011kd, Domazet:2012tw, Rodrigues:2014xka, Koch:2014joa}. In order to be clear in respect to our objectives, it is not the purpose of this work to consider RG group effects to $f(R)$ gravity or to ST gravity (e.g., \cite{Codello:2007bd, Benedetti:2013nya, Percacci:2015wwa, Shapiro:2015ova}), but to compare the RG improved Einstein-Hilbert action to ST gravity at large scales.

The field equations and the phenomenology derived from the RG approach described above depends on a certain scale which we call $\mu$. The values of $G$ and $\Lambda$ depend on this scale (analogous to what happens in scattering experiments within QED, where the effective fine structure constant changes its value depending on a certain scale, which in this case is given by the momenta of the scattering particles). Hence, any proposal in this line must address three issues: i) the relation between $G$ and $\mu$, or equivalently provide a $\beta$-function for $G$ ($\beta_G$), ii) a $\beta$-function for $\Lambda$ ($\beta_\Lambda$), iii) the relation between $\mu$ and other physically relevant quantities (i.e., a scale setting procedure \cite{Babic:2004ev}). 

As argued in \cite{Shapiro:2004ch,Babic:2004ev, Reuter:2003ca}, these three pieces of information cannot be set arbitrarily, since the classical dynamics (the energy-momentum tensor conservation in particular) imply a relation between them. Actually, and this is stressed in this work, from the knowledge of two of them, the field equations can be used to derive the third one. A particular form of standard ST gravity can be found if $\beta_\Lambda$ and $\beta_G$ are assumed to be known at the action level (equivalently, if $\Lambda(\mu)$ and $G(\mu)$ are known functions at the action level, i.e. if $\Lambda(\mu)$ and $G(\mu)$ are fixed for any matter content). In this case, $\mu$ is a scalar field whose solution is derived from the field equations. Assuming that certain function has an inverse, it is natural to obtain an $f(R)$ theory from this procedure \cite{Koch:2010nn,Hindmarsh:2012rc, Cai:2011kd, Domazet:2012tw}, and $\mu$ can be interpreted as an auxiliary field \cite{Pani:2013qfa}.

Another possible approach, which we explore here in detail, considers that $\beta_G$ is fixed at the action level, $\mu$ is chosen from  physical considerations, and the relation between $\Lambda$ and $\mu$ is derived from the field equations. This implies that the $\beta$-function associated with $\Lambda$ leads not to a universal $\Lambda(\mu)$ function, and that it is sensitive to the matter distribution and boundary conditions.  This is the main line that is developed here.\footnote{{\it A priori}, one may  consider the case in which $\beta_G$ is the $\beta$-function sensitive to the matter distribution. However, since $\Lambda$ is a peculiar coupling constant, it seems more natural to assume that $G(\mu)$ is the function to be fixed as a universal function from QFT in curved spacetime or quantum gravity considerations, see eq. (\ref{eq:Gmu}) in particular.} Examples of this approach include Refs. \cite{Reuter:2003ca, Domazet:2010bk, Shapiro:2004ch, Rodrigues:2009vf}. Nevertheless, these references  state that $\Lambda$ is either negligible for the phenomenology considered, or that it is choosen such that it is compatible with the field equations. In this work we consider in great detail the approach of {\it deriving} the relation between $\Lambda$ and $\mu$ from the field equations. Clearly, this framework is analogous to the extended ST picture described in the beginning of the Introduction, where the potential is system dependent. We also clarify in detail that the assumption that either $G$ or $\Lambda$ are external scalar fields, as used in \cite{Reuter:2003ca, Rodrigues:2009vf}, is neither necessary nor it is the main difference in regard to ST gravity (provided that certain usual condition is met, see also \cite{Chauvineau:2015cha}).

\bigskip

This work is organised as follows: in the next section we review  and present  action approaches to the large scale RG effects in gravity, in particular we present a new class of covariant scale setting procedures in Subsec.~\ref{sec:scalesettingproposal}. Section \ref{sec:Lsol} is devoted to present and exemplify methods for the derivation of $\Lambda$ (from the knowledge of $G(\mu)$ and a scale setting for $\mu$). Section \ref{sec:STcomparison}  details the extended ST picture with system dependent potential and compares it with the Einstein-Hilbert action with RG effects. Our conclusions are presented in section~\ref{sec:conclusions}.

\section{Effective action approaches to the Renormalization Group effects at large scales} \label{sec:RGactions}

\subsection{The RG improved action with external \texorpdfstring{$G$}{G} and \texorpdfstring{$\Lambda$}{L}}

In Refs. \cite{Reuter:2003ca, Reuter:2004nx, Reuter:2004nv, Shapiro:2004ch, Reuter:2007de, Rodrigues:2009vf}, see also \cite{Koch:2010nn, Cai:2011kd, Domazet:2012tw,  Rodrigues:2012qm, Rodrigues:2014xka, Koch:2014joa}, it is argued in favour of the following action capable of enclosing the large scale Renormalization Group effects for gravity
\begin{equation}
    S[g] = \frac{1}{16 \pi} \int \frac{R - 2 \Lambda}{G} \sqrt{-g}   d^4x. \label{eq:RGexter}
\end{equation}
In the above, $G$ and $\Lambda$ are not true constants, their values depend on a certain RG scale, which we label $\mu$ ($k$ is another commonly used notation). On the other hand, neither $G$ nor $\Lambda$ are standard classical scalar fields. They behave as scalars under coordinate transformations, but, as it is stressed in the above references, $G$ and $\Lambda$ are external fields. In particular, Refs. \cite{Reuter:2003ca, Reuter:2007de} stress the external character of $G$ and $\Lambda$ as the main difference between the action (\ref{eq:RGexter}) and usual ST theories.\footnote{In these references, sometimes $G$ and $\Lambda$ are also referred as ``background fields''.} 

Although the above action is expected to describe certain RG effects, which will be detailed shortly, it should not be seen as capable of a complete description of either quantum gravity or full QFT in curved space-time. From the perspective of QFT in curved space-time, the higher derivative terms, necessary for the perturbative renormalizability, are missing, hence the above action can  be seen as an approximation for gravity (with RG effects) at scales much larger than the Planck one and for weak fields (i.e., far from black hole horizons or the big bang). From the perspective of the Asymptotic Safety program, a similar conclusion also holds. The action above  can describe quantum effects within the Einstein-Hilbert truncation, but this truncation is severe and not  supposed to be sufficient to yield fully coherent quantum gravity picture \cite{Niedermaier:2006wt, Becker:2014qya}.\footnote{Black holes in the context of Asymptotic Safety program and the Einstein-Hilbert truncation were studied in some refs, see e.g. Ref. \cite{Bonanno:2000ep, Bonanno:2006eu, Koch:2014cqa}. Corrections to the General Relativity dynamics  are typically studied within the RG improved solutions approach, which does not seem to be the most realistic approach, but it may open the way for future developments. Further comments on black holes and the RG improved action approach can be found in \cite{Koch:2014cqa}.} Nonetheless, since the emphasis in this work is on the General Relativity deviations at large scales, even if corrections beyond he Einstein-Hilbert truncation are relevant for the large scale RG flow of $G$ and $\Lambda$, their direct influence may be irrelevant for the dynamics at large scales. In conclusion, regardless on the fundamental approach towards the RG and gravity, the above action should be seen as an approximation for gravity with RG corrections at large scales.

By labelling $G$ and $\Lambda$ as external scalar fields, it is meant that, in order to derive the field equations from action (\ref{eq:RGexter}), one should only consider variations with respect to the metric $g_{\mu\nu }$, and not with respect to the scalar quantities $G$ and $\Lambda$ (that is why action (\ref{eq:RGexter}) was labeled a functional of $g_{\mu\nu }$ alone) (on external scalar fields, see also \cite{Chauvineau:2015cha, Boehmer:2013ss}). Since the running of $G$ and $\Lambda$ are rooted on non-classical arguments, it is appealing to consider them external fields in the context of an effective classical action.

All the field equations derived from action (\ref{eq:RGexter}), added to matter fields,  read 
\begin{equation}
    \label{eq:fieldext}
    {\cal G}_{\alpha \beta} + \Lambda g_{\alpha \beta} = {8 \pi G} T_{\alpha \beta},
\end{equation}
where 
\begin{equation}
        {\cal G}_{\alpha \beta} \equiv G_{\alpha \beta} +  G \Box G^{-1} g_{\alpha \beta} - G \nabla_\alpha \nabla_\beta G^{-1},
\end{equation}
$\Box \equiv g^{\alpha \beta} \nabla_\alpha \nabla_\beta$, and $\nabla_\alpha$ is the usual covariant derivative. These field equations are not sufficient for deriving $g_{\mu\nu }$, $G$ and $\Lambda$ as spacetime functions, but they impose restrictions on these fields. In particular,  $G$ and $\Lambda$ must be such that a certain classical consistency equation is satisfied \cite{Reuter:2003ca}. This equation is derived below.

By assuming that the energy-momentum tensor $T_{\alpha \beta}$ is conserved,\footnote{This conservation can be derived if $T_{\alpha \beta}$ is obtained from a scalar action, but only if this action includes all the matter fields and do not depend on the external fields $G$ and $\Lambda$ \cite{Reuter:2003ca, Chauvineau:2015cha}.} which is also in accordance with Refs. \cite{Reuter:2003ca, Reuter:2004nx, Reuter:2004nv, Shapiro:2004ch, Reuter:2007de, Rodrigues:2009vf} approach, one derives that
\begin{eqnarray}
 \nabla_\alpha \Lambda+ \nabla_\alpha (G \Box G^{-1}) - \nabla^\beta ( G \nabla_\alpha \nabla_\beta G^{-1}) &=&  {8 \pi} (\nabla^\beta G) T_{\alpha \beta}, \nonumber \\ 
 &=& G^{-1} (\nabla^\beta G) ({\cal G}_{\alpha \beta} + \Lambda g_{\alpha \beta}),
\end{eqnarray}
hence
\begin{equation}
    \label{eq:aux1}
    \nabla_\alpha \Lambda + \Lambda G \nabla_\alpha G^{-1} + G (\nabla^\beta G^{-1}) G_{\alpha \beta}  + G \nabla_\alpha \Box G^{-1} - G \nabla_\beta \nabla_\alpha \nabla^\beta G^{-1} = 0.
\end{equation}
For any vector field $A^\mu$, one has $[\nabla_\alpha, \nabla_\beta] A^\beta = A^\sigma R^\beta_{\; \sigma \alpha \beta} = - A^\sigma R_{\sigma \alpha}$, hence
\begin{equation}
    G \nabla_\alpha \Box G^{-1} - G \nabla_\beta \nabla_\alpha \nabla^\beta G^{-1} = - G (\nabla^\sigma G^{-1})R_{\sigma \alpha}.
\end{equation}
Finally, by inserting the above result into Eq.~(\ref{eq:aux1}),
\begin{equation}
    \label{eq:consist}
    \nabla_\alpha \left( \frac{\Lambda}G \right) = \frac 12 R \nabla_\alpha G^{-1}. 
\end{equation}
In conclusion, the field equations (\ref{eq:fieldext}) together with energy-momentum  conservation implies Eq.~(\ref{eq:consist}). The equation above, in that compact form, can also be found in a number of works \cite{Koch:2010nn,Cai:2011kd, Rodrigues:2012qm}. Reference \cite{Reuter:2003ca} presents an equation that is equivalent to the above one, and calls it consistency equation, since it imposes a necessary condition on  $G$ and $\Lambda$ in addition to any quantum conditions.\footnote{This condition is a direct consequence of the RG improved action approach, a different condition appears in the RG improved field equations approach, and no additional condition is derived from the RG improved solutions approach. For further details on this classification, see \cite{Reuter:2003ca}.}  We use the same nomenclature.

\subsection{The RG improved action without external scalar fields}\label{sec:RGwihoutexternal}

The use of external fields is not a problem by itself for classical dynamics, but it is an inconvenience since it obscures the classical dynamics and the relation between this and other gravitational models. Actually, if the energy-momentum tensor is assumed to be conserved, it is not hard to realize that the action (\ref{eq:RGexter}) can be re-expressed without external fields (in case certain condition is met, details below). 

Consider the action
\begin{equation}
    \label{eq:RGNOexter}
    S[g, \mu] = \frac{1}{16 \pi} \int \frac{R - 2 \Lambda(\mu)}{G(\mu)} \sqrt{-g} \,  d^4x.
\end{equation}
To our knowledge, in the context of the RG application to gravity, this action with that functional dependence was first considered in Ref. \cite{Koch:2010nn}, and in particular it was also used in Refs. \cite{Rodrigues:2012qm, Rodrigues:2014xka}. By varying  the above action with respect to the metric one finds the field equations (\ref{eq:fieldext}), whilst the variation of the above with respect to the scalar $\mu$ leads to
\begin{equation}
    \label{eq:consistmu}
    \frac{d}{d\mu} \left( \frac{\Lambda}G \right) = \frac 12 R \frac{d}{d\mu}G^{-1}. 
\end{equation}
Equations (\ref{eq:consist}, \ref{eq:consistmu}) are almost identical. It is assumed that $\mu$ is a field and that both $G$ and $\Lambda$ are functions of $\mu$, hence  it is always possible to write $\frac{\partial}{\partial x^\alpha} G= \frac{\partial \mu }{\partial x^\alpha} \frac{d}{d\mu}G$, and therefore  Eq.~(\ref{eq:consist}) can be written as
\begin{equation}
     \label{eq:consist2}
    \nabla_\alpha \mu \left[ \frac{d}{d\mu} \left( \frac{\Lambda}G \right) - \frac 12 R \frac{d}{d\mu}G^{-1}\right] =0.
\end{equation}

In conclusion, if Eq.~(\ref{eq:consistmu}) is satisfied, necessarily Eq.~(\ref{eq:consist}) is also true, and hence $\nabla^\alpha T_{\alpha \beta} =0 $. On the other hand, in principle it is possible that  Eq.~(\ref{eq:consist}),  Eq.~(\ref{eq:consist2}) and $\nabla^\alpha T_{\alpha \beta} =0$ are satisfied, but Eq.~(\ref{eq:consistmu}) is not. This may only happen in the case that $\nabla_\alpha \mu =0$ in some spacetime region. This possibility is developed in Ref. \cite{Chauvineau:2015cha}.

The latter subtlety is the only possible dynamical difference between the action (\ref{eq:RGexter}) with energy-momentum tensor conservation and the action (\ref{eq:RGNOexter}). Such subtlety was not used or discussed in the previous references on the subject. In conclusion, the main difference between RG-induced modifications to the Einstein-Hilbert action and Scalar-Tensor gravity does not reside in the use of external scalar fields, unless one of the following nonstandard possibilities is considered: either  $\nabla^\alpha T_{\alpha \beta} \not=0$ or, at certain spacetime regions, $\nabla_\alpha \mu = 0$.

\subsection{Scale setting at the action level} \label{sec:dynamicallycomplete}

The action (\ref{eq:RGNOexter}) can be seen as an improvement over the action (\ref{eq:RGexter}),\footnote{Their dynamical content is the same assuming  $\nabla_\alpha T^{\alpha \beta}=0$ and $\nabla_\alpha \mu \not=0$, whilst  action (\ref{eq:RGNOexter}) does not depend on external fields.} nevertheless there is still a piece of information that was not explicitly added to the action, the scale setting.\footnote{In Ref. \cite{Reuter:2003ca}, the authors argue in favor of the scale setting (or the cut-off specification) at the action level, which lead them, for an arbitrary scale $\mu$, to action (\ref{eq:RGexter}). In that action, $G$ can be expressed as a spacetime function since the scale (independently on its nature) should be in the end expressible as a spacetime function. Hence, in action (\ref{eq:RGNOexter}),  $\mu$ is an arbitrary scale and one should specify its meaning.} In diverse approaches,  the scale setting is considered as an additional equation alongside the field equations, e.g. \cite{Reuter:2003ca, Reuter:2004nv, Rodrigues:2009vf, Rodrigues:2012qm, Sola:2013fka, Lima:2014hia}. Nevertheless, in this approach of the RG improved action, it would be natural, and desirable, to provide every piece of necessary information in the action.

Albeit $\mu$ in the end should be expressed as a spacetime function, at a more fundamental level there should be some relation of this scale with other fields of physical relevance, hence in general the scale setting should have the form,
\begin{equation}
	\mu = f(g,\Psi),
\end{equation}
that is, $\mu$ should be expressed, in general, as a function of the metric and matter fields, the latter we do not specify the nature and collectively label them $\Psi$. It is not explicit in the above notation, but the function $f$ may also depend on the derivatives of these fields.

Actually  action (\ref{eq:RGNOexter})  contains a hidden scale setting, which comes from Eq.~(\ref{eq:consistmu}), as detailed in Refs. \cite{Koch:2010nn,Domazet:2012tw, Hindmarsh:2012rc}. Namely, Eq.~(\ref{eq:consistmu}) states that $R$ is equal to a function of $\mu$. Then, if this function has an inverse,  $\mu$ can be written as a function of $R$, and  action (\ref{eq:RGNOexter}) becomes equivalent to a particular case of $f(R)$ action. 

This structure may seem as  unexpected from the RG perspective, since the scale setting does not come from an off-shell relation derived from RG arguments, but it is derived from the field equations. The consistency equation imposes a link between three RG relations, namely: $i$) the relation between $G$ and $\mu$, $ii$) the one between $\Lambda$ and $\mu$, and $iii$) the scale setting. Hence, if one starts from the assumption that the relations $i$ and $ii$ are fixed off-shell (i.e., $G(\mu)$ and $\Lambda(\mu)$ are given functions fixed at the action level), then the third one is fixed from the field equations. Here we shall consider another route that was not previously explored in generality,\footnote{Particular aspects of this route were explored in Refs. \cite{Reuter:2003ca, Rodrigues:2012qm, Rodrigues:2014xka}.} namely considering that the relations $i$ and $iii$ are given at the action level, while $ii$ is determined from the field equations. In this case, $\beta_G$ and $\mu$ have off-shell expressions (and hence they are necessarily system independent), but  the gravitational parameter $\Lambda$ is associated with an on-shell $\beta$-function (that is, a $\beta$-function that is not directly derived from pure QFT or quantum gravity expectations, but that uses the classical field equations). This implies that the form of the $\beta_\Lambda$ (and hence the relation between $\Lambda$ and $\mu$) depends in general on properties associated with other fields (including their boundary conditions), that is, depending on other fields properties at a given spacetime region, one may have either, say,  $\Lambda \propto \mu^2$ or $\Lambda \propto \ln \mu$. As far as we know, this possible change on $\beta_\Lambda$ depending on the system has not been systematically explored, but it is a natural consequence once one uses the field equations in order to infer $\beta$-functions. Indeed, this is not a usual feature for other forces, but gravity is significantly different from the other forces in diverse ways. It is a  route that seems viable and hence needs to be explored.  

We introduce the following notation: let $F(\mu)$ be the usual function that is defined off-shell (i.e., at the action level), and let  $F\{\mu\}$ be  a function that is not fixed at the action level, hence, in general, its expression as a $\mu$ function depends on the boundary conditions of the system, hence it is system-dependent. If, for a given system, one  derives the $F\{\mu\}$ expression and, for the same system, re-inserts it in the action in the form of an off-shell function $F(\mu)$, then it is expected that the field solutions will be exactly the same. This is detailed in section~\ref{sec:STcomparison}. Nonetheless, $F\{\mu\}$ typically has nontrivial dependence on $\mu$ (it may be non-analytical) and for different systems it will change, while $F(\mu)$ is fixed for any system. 

 One way of achieving the scale setting at the action level is through the use of a Lagrange multiplier as follows,
\begin{equation}
    \label{eq:RGNOexterConst}
    S[g,\mu,\lambda,\Psi] =  \int \left[ \frac{R - 2 \Lambda\{\mu\}}{16 \pi G(\mu)}  + \lambda \left( \mu - f(g,\Psi)\right)\right] \sqrt{-g} \,  d^4x + S_{\mbox{\tiny matter}}[g,\Psi].
\end{equation}
 The form of the field equations that are derived from this action do not depend on the use  of $\Lambda\{\mu\}$ or $\Lambda(\mu)$; but the solutions are different in general, since $\Lambda(\mu)$ is fixed to be the same for every system, while $\Lambda\{\mu\}$ is derived for each particular system. Explicit examples on $\Lambda \{ \mu\}$ derivations are shown in section~\ref{sec:Lsol}. General properties and the consistency of the action (\ref{eq:RGNOexterConst}) are detailed in section~\ref{sec:STcomparison}. Comparing with standard ST gravity, the equality $\mu = f(g,\Psi)$ imposes an additional restriction, but the relation between $\Lambda$ and $\mu$  is left unspecified at the action level.

If it is possible to express $\mu$ as a function of $G$,  then action (\ref{eq:RGNOexterConst}) is equivalent to 
\begin{equation}
    \label{eq:RGNOexterConstNOmu}
    S[g,G,\lambda,\Psi] = \int \left[ \frac{R - 2 \Lambda\{G\}}{16 \pi G}  + \lambda \left( G - F(g,\Psi)\right)\right] \sqrt{-g} \,  d^4x + S_{\mbox{\tiny matter}}[g,\Psi].
\end{equation}
It is straightforward to check that the actions given by Eqs.~(\ref{eq:RGNOexterConst}, \ref{eq:RGNOexterConstNOmu}) lead to  equivalent field equations whenever one can invert $G(\mu)$ to express $\mu(G)$. This form will be useful to the comparison with the Brans-Dicke action in section \ref{sec:STcomparisonOntheactions}.

Another relevant way to express action (\ref{eq:RGNOexterConst}) comes from the elimination of the scalar $\mu$, which leads to the action,
\begin{equation} \label{eq:RGNOexterConstNOscalar}
		\bar S[g,\Psi]=\int \frac{R-2\Lambda \{f(g,\Psi)\}}{16\pi G(f(g,\Psi))}\sqrt{-g}\, d^4x+S_{\mbox{\tiny matter}}[g,\Psi].
\end{equation}
For completeness,  Appendix \ref{app:elimination} shows the equivalence between actions $S$ and $\bar S$. The above form shows that the RG improved Einstein-Hilbert action can be seen as a pure metric gravitational theory with a nonstandard coupling to matter.  

If $\partial f/ \partial \Psi \not=0$, actions (\ref{eq:RGNOexterConst}, \ref{eq:RGNOexterConstNOmu}, \ref{eq:RGNOexterConstNOscalar}) will  in general spoil energy-momentum tensor conservation, but as it will be seen in the next section, there is a particular class of scale settings (i.e., functions $f$) with $\partial f/ \partial \Psi \not=0$  that preserves energy-momentum conservation.

In order to develop particular examples, a particular and natural expression for $G(\mu)$ will be necessary. The following  was derived from different approaches (see e.g., \cite{Reuter:2003ca,  Shapiro:2004ch, Bauer:2005rpa}), and it can be inverted,\footnote{Here $\mu$ is written as a dimensionless quantity, for the dimensionful case one should replace $\mu$ by the fraction $\mu/\mu_0$, where $\mu_0$ is a constant.} 
\begin{equation}
    \label{eq:Gmu}
    G(\mu) = \frac{G_0}{1 + 2 \nu \ln \mu},
\end{equation}
where $\nu$ is a small dimensionless constant. General Relativity corresponds to the $\nu=0$ case.

\subsection{A covariant scale setting proposal and its application for relativistic fluids} \label{sec:scalesettingproposal}

In the context of stationary, slow velocity and weak field systems, some of us have introduced in previous works the scale setting 
\begin{equation}
    \label{eq:muRGGR}
    \mu = f\left( \frac{\Phi_N}{\Phi_0} \right),
\end{equation}
where $\Phi_N$ is the Newtonian potential, defined by $\nabla^2 \Phi_N = 4 \pi G_0 \rho$ (where $\rho$ is a matter density profile) with $\Phi(r \rightarrow \infty) = 0$, $\Phi_0$ is some constant and $\mu$ is the RG scale written in dimensionless form. Such setting has achieved  interesting phenomenological achievements, and we speculated on its possible connection to dark matter \cite{Rodrigues:2009vf, Rodrigues:2011cq, Fabris:2012wg, Rodrigues:2012qm, Rodrigues:2012wk, Rodrigues:2014xka, deOliveira:2015cja}.

In order to both evaluate such proposal in other contexts, and to compare it to ST gravity, it is necessary to express the relation (\ref{eq:muRGGR}) in a covariant way. It is well known that in the weak field limit General Relativity provides a straight relation between $\Phi_N$ and $g_{00}$, namely $g_{00} = -1 - 2 \Phi_N$. One should note that there is no nontrivial scalar quantity that can be generated by the metric alone. If derivatives are considered, then the simplest scalar is $R$, but this scalar has no straight relation with\footnote{In the weak field limit it  has a straight relation with $\nabla^2 \Phi_N$.} $\Phi_N$. The Newtonian potential $\Phi_N$ is capable of encoding all the relevant information of gravitational systems within weak fields and small velocities. In particular, for a massive particle at $r=0$, the Newtonian potential reads $\Phi_N = - G_0 M /r$, while the Ricci scalar $R$ is simply a constant (equal to $4 \Lambda$) for any $r>0$.

One way of generating the correspondence given by (\ref{eq:muRGGR}) is to consider additional vectorial quantities  in order to single out a preferred frame (these vectorial quantities may or may not be written as the gradient of scalar quantities). It should be stressed that a reference frame that is at rest with respect to a certain system is not, in general, at rest in the preferred frame given by the vectorial quantities. Therefore, if this route is pursued, the relation (\ref{eq:muRGGR}) would depend on the difference between the rest frame of the system and the preferred frame given by the additional vector fields. 

A standard approach to covariantize a quantity whose expression is known in a certain rest frame is to construct scalars by using the 4-velocity $U^\alpha$. This approach circumvents the issue indicated in the previous paragraph, since the vectorial quantity considered is the 4-velocity, hence the ``preferred" frame given by the vectorial structure and the rest frame are the same by definition. For instance, if the quantity $j_{00}$ is known in the frame in which $U^i \approx 0$, one can compute the value of the scalar $U^\alpha U^\beta j_{\alpha \beta} \approx U^0 U^0 j_{00}$ in that frame, and that value will hold in any frame. Nevertheless, for the case being considered, the corresponding scalar quantity is just a constant since $U^\alpha U^\beta g_{\alpha \beta} = - 1$, hence this approach in this form cannot be used to covariantize the ansatz in Eq.~(\ref{eq:muRGGR}).

For either a particle or a fluid whose four-velocity is denoted by $U^\alpha$, we consider the following covariant extension for Eq.~(\ref{eq:muRGGR}),
\begin{equation}
    \label{eq:mufUUh}
    \mu = f\left( { U^\alpha U^\beta h_{\alpha \beta}} \right),
\end{equation}
where $h_{\alpha \beta} \equiv  g_{\alpha \beta} - \gamma_{\alpha \beta}$ and $\gamma_{\alpha \beta}$ is a rank-two tensor. In order to  disclose Eq.~(\ref{eq:muRGGR}) from the above ansatz, one should look for solutions in which $\gamma_{\alpha \beta}$ is a Minkowski metric. 

The appearance of an additional rank two tensor in the context of the Renormalization Group application to gravity is not a novelty (see, e.g. \cite{Reuter:2008wj,Manrique:2010mq, Manrique:2010am, Becker:2014qya}).

Considering this proposal the corresponding action in the form (\ref{eq:RGNOexterConst}) coupled to a relativistic fluid is
\begin{equation}
    \label{eq:RGNOexterConstUUh}
    S[g,\mu,\lambda, \gamma, U,\Psi] =  \int \left[ \frac{R - 2 \Lambda\{\mu\}}{16 \pi G(\mu)}  + \lambda \left( \mu - f\left( { U^\alpha U^\beta h_{\alpha \beta}} \right) \right) \right] \sqrt{-g} \,  d^4x + S_{\mbox{\tiny matter}}[g, U, \Psi].
\end{equation}
Depending on the fluid description, $U^\alpha$ may or may not be a fundamental field. We consider the description presented in Ref.~\cite{1972JMP....13.1451R}, in which $U^\alpha$ is a fundamental field and the fundamental thermodynamical variables are the rest mass density $n$ and the rest specific entropy $s$,
\begin{equation}   
     \label{eq:Sfluid}
    S_{\mbox{\tiny matter}} = \int \left[-\rho(n,s) + \eta_1 (1 + U^\alpha U_\alpha) + \eta_2 \nabla_\alpha (n U^\alpha) + \eta_3 U^\alpha  \nabla_\alpha X +  \eta_4 U^\alpha  \nabla_\alpha s \right]\sqrt{-g}\,d^4x.  
\end{equation}
In the above $S_{\mbox{\tiny matter}}  = S_{\mbox{\tiny matter}} [g, U, n,s, \eta_p,X]$, $\eta_p$ stands for the Lagrange multipliers $\eta_1,..,\eta_4$ and $\rho(n,s)$ is the energy density. The quantity $X$ express the ``particle identity'', and is important for the description of fluids with rotational flow \cite{1972JMP....13.1451R}. 

In the following we show  the influence of the term $\lambda (\mu - f(U^\alpha U^\beta h_{\alpha \beta}))$ to the perfect fluid physics. From the action $S$ variation with respect to $n$ and $U^\alpha$, one derives respectively,
\begin{equation}
	\partial_n \rho + U^\alpha \partial_\alpha \eta_2 =0, \label{eq:Fluidn}
\end{equation}
\begin{equation}
	- 2 \lambda f'U^\beta h_{\alpha \beta} + 2 \eta_1 U_\alpha - n \partial_\alpha \eta_2 + \eta_3 \partial_\alpha X + \eta_4 \partial_\alpha s =0, \label{eq:FluidU}
\end{equation}
where $f'$ is the derivative of the function $f$. Therefore, contracting Eq.~(\ref{eq:FluidU}) with $U^\alpha$ and using both Eq.~(\ref{eq:Fluidn}) and the constraints associated with $\eta_3$ and $\eta_4$, 
\begin{equation}
	\eta_1 = - \lambda f'U^\alpha U^\beta h_{\alpha \beta} + \frac 12 n \partial_n \rho.
\end{equation}

With the above results, we proceed to derive the $S$ variation with respect to the metric, namely,
\begin{eqnarray}
	{\cal G}_{\alpha \beta} + \Lambda g_{\alpha \beta} &=& 16 \pi G \left[ \left(\eta_1 - \lambda f'  \right) U_\alpha U_\beta  +\frac 12 g_{\alpha \beta} (n \partial_n \rho - \rho)\right]	 \nonumber\\[.1in]
	&=& 8 \pi G \left[(P + \rho - 2 (U^\kappa U^\sigma h_{\kappa \sigma}+1) \lambda f') U_\alpha U_\beta + P g_{\alpha \beta}  \right],
\end{eqnarray}
where it was introduced $P \equiv  n \partial_n \rho - \rho$ \cite{1972JMP....13.1451R}. Clearly the right hand side of the above equation only corresponds to the usual perfect fluid energy-momentum tensor if the term proportional to $\lambda f'$ is zero. Indeed, from the variation of $S$ with respect to $\gamma_{\alpha \beta}$,
\begin{equation}
	\lambda f'U^\alpha U^\beta = 0,
\end{equation}
and hence either $\lambda$ or $f'$ are zero. The last option leads to standard General Relativity, since it implies that $\mu$ and $G$ are constants. The case of interest  corresponds to the solution $\lambda =0$. This case is compatible with the running of $G$ and $\Lambda$ and does not directly change any fundamental property of the perfect fluid. In particular, one can  check that the energy-momentum tensor has the standard  form, and that it is conserved (due to its own equations of motion). This  is expected since only the terms proportional to $\lambda$ may change any of the fluid properties. Moreover, since $\lambda$ is zero at the level of the field equations, it is easy to verify that Eq.~(\ref{eq:consistmu}) also follows from the variation of $S$ with respect to $\mu$.

\section{Procedures for deriving  \texorpdfstring{$\Lambda$}{L} solutions} \label{sec:Lsol}

The main purposes of this section are to show, with particular examples, how the solution for $\Lambda$ can be derived and how this solution can change depending on the system properties. These solutions can be expressed as perturbations on a dimensionless RG parameter, which we call $\nu$, and is such that for $\nu=0$ one recovers General Relativity, see Eq. (\ref{eq:Gmu}).

There are some ways of deriving $\Lambda$. For some systems it is possible to first derive the metric solution and then derive the $\Lambda$ solution \cite{Rodrigues:2015rya}, but since the focus here is on $\Lambda$ and not on the metric solutions, the most straightforward way is to use directly the consistency equation (\ref{eq:consist}).\footnote{Note that in general it is not possible to perform an integration on $\mu$ to solve Eq.~(\ref{eq:consistmu}), since the metric solutions should depend on $\mu$, and this relation is unknown  {\it a priori}. An  exception of the latter observation is the case in which $\mu$ is fixed such that $\mu = f(R)$, in this case it is always possible to integrate Eq.~(\ref{eq:consistmu}) on $\mu$ (assuming that the function $f$ has an inverse), even before knowing the metric solutions. Albeit technically easier, this identification is not favoured, see section \ref{sec:scalesettingproposal}.} However, since in many cases the matter content is known but the metric solutions are not, it is sometimes more convenient to eliminate the $R$ dependence in favour of a $T_{\alpha \beta}$ dependence. The trace of Eq.~(\ref{eq:fieldext}) reads
\begin{equation}
    R = 3 G \Box G^{-1} + 4 \Lambda - {8 \pi G} T,
\end{equation}
where $T \equiv g^{\alpha \beta} T_{\alpha \beta}$. Hence Eq.~(\ref{eq:consist}) is equivalent to
\begin{equation}
    \label{eq:consistrho}
    \nabla_\alpha \Lambda = \left( \frac 32 G \Box G^{-1} + \Lambda - {4 \pi G} T\right) G \nabla_\alpha G^{-1}.
\end{equation}
The above relation is exact, no approximations were done. This expression may not seem as a significant improvement for the case in which metric solutions are unknown, since the $R$ dependence was eliminated, but the term $ G \Box G^{-1} $ was introduced, which also depends on the metric. However, for the perturbative picture, Eq.~(\ref{eq:consistrho}) is an improvement since the metric only appears in a term that is at least of $\nu$ order.

\subsection{Renormalization Group perturbation about exact General Relativity solution}

\subsubsection{Vacuum}  \label{sec:RGpertGRexactvacum}

We will consider the expression for $G(\mu)$ as given by\footnote{To be more precise, for this subsection we do not need the full details of Eq. (\ref{eq:Gmu}), we only use that  $G|_{\nu=0} = G_0$ and that $G^{-1}$ depends at most on the first order of $\nu$.} Eq.~(\ref{eq:Gmu}) and the following $\Lambda$ expansion
\begin{equation}
    \Lambda = \Lambda_0 + \nu \Lambda_1 + O(\nu^2),
\end{equation}
where $\Lambda_0$ is a constant, since, by construction, for $\nu=0$ one needs to recover General Relativity.  Also, the last condition implies that the metric solution and the energy-momentum tensor can be written as
\begin{eqnarray}
    g_{\alpha \beta} &=& \stackrel{(0)}g_{\alpha \beta} + \nu \stackrel{(1)}g_{\alpha \beta} + O(\nu^2), \label{eq:perturbRGg}\\[.1in]
    T_{\alpha \beta} &=& \stackrel{(0)}T_{\alpha \beta} + \nu \stackrel{(1)}T_{\alpha \beta} + O(\nu^2), \label{eq:perturbRGT}
\end{eqnarray}
where $\stackrel{(0)}g_{\alpha \beta}$ is the General Relativity solution with cosmological constant $\Lambda_0$ and energy-momentum tensor $\stackrel{(0)}T_{\alpha \beta}$.

At zeroth order on $\nu$, Eq.~(\ref{eq:consistrho}) is trivially satisfied, since both $G$ and $\Lambda$ are constants. Up to the first order on the RG effects,  with $\stackrel{(0)}T \equiv \stackrel{(0)}{g}\,\!\!\!^{\alpha \beta} \stackrel{(0)}T_{\alpha \beta}$, it reads
\begin{equation}
    \label{eq:Lambda1}
    \nu \nabla_\alpha \Lambda_1 = \left(  \Lambda_0 - {4 \pi G_0} \stackrel{(0)}T \right)  \, G_0 \nabla_\alpha  G^{-1} + O(\nu^2).
\end{equation}
For $\stackrel{(0)}T=0$, and using the boundary condition $\Lambda(G=G_0) = \Lambda_0$, the above equation has a simple solution which reads
\begin{equation}
    \label{eq:lambdaT0null}
    \Lambda =  \Lambda_0 G_0 G^{-1} + O(\nu^2).
\end{equation}
The above result does not depend on any particular symmetry, hence it is quite general. On the other hand, this result is only guaranteed to hold if $\stackrel{(0)}T=0$. In general the relation between $\Lambda$ and either $G$ or $\mu$ is system dependent. This should become clear with the particular cases that are presented afterwards.

The result (\ref{eq:lambdaT0null}) can also be written as,
\begin{equation}
	\Lambda G = \mbox{constant} + O(\nu^2). \label{eq:LGconstant}
\end{equation} 
This presents a necessary condition which the {\it complete} RG flows of $G$ and $\Lambda$ need to satisfy in vacuum in order to yield a consistent (semi-)classical picture. The use of ``complete'' is to stress that all the relevant physical issues related with the RG flow need to be taken into account, including in particular backreaction effects, RG flows corrections that go beyond the Einstein-Hilbert truncation and the existence or not of infrared (IR) fixed points. For instance, the $G$ and $\Lambda$ flows presented in Refs.~\cite{Reuter:1996cp, Bonanno:2001xi}, satisfy the condition above close to the UV fixed point, but not in a certain long wave length limit, which assumes no IR fixed point.  See also Refs.~\cite{Bonanno:2001hi,Bentivegna:2003rr}, where the existence of an IR fixed point is considered. Such uncertainties at the fundamental level (which may prove impossible to compute directly) need not to be fully addressed directly since the result from Eqs.~(\ref{eq:lambdaT0null}, \ref{eq:LGconstant}) allows to derive the RG flow of one of them, say $\Lambda$, from the knowledge of the other one which is assumed to be known with higher accuracy, say $G$.


\subsubsection{Sphere of constant density} \label{sec:RGpertGRexactInterior}

Before proceeding towards another perturbative scheme, we present an example with $\Lambda_0=0$ and $\stackrel{(0)}T\not=0$. Equation (\ref{eq:Lambda1}) alone shows that a zero value for $\Lambda_0$ does not imply that $\Lambda$ will necessarily be null, since if matter is present it is possible to have $\Lambda_0=0$ and $\Lambda \not=0$. 

We consider the simplest exact General Relativity solution with spherical symmetry, static and non negligible energy density $\rho$ (the ``interior Schwarschild solution''). Let $\rho$ be a positive constant inside the radius $R$, and zero outside. Considering no cosmological constant, the line element solution is well known within General Relativity and it reads (e.g., \cite{9780521829519})
\begin{equation}
    ds^2 = -A(r) dt^2 + B(r) dr^2 + r^2 d\Omega^2, \label{eq:linespherical}
\end{equation}
with
\begin{eqnarray}
    && A(r) = \left ( \frac 32 \sqrt{1 - 2 \frac M{R}} - \frac 12 \sqrt{1 - 2 \frac M{R^3}r^2}\right)^2, \label{eq:A(r)}\\
    && B(r) = \frac{1}{1 - 2 \frac{M}{R^3}r^2},
\end{eqnarray}
where $M$ is a constant. For the derivation of the above solution, the boundary conditions at $R$ are such that the metric is continuous and differentiable at any point. In order for this mass distribution to be static, the fluid of constant density must have a nontrivial pressure. Indeed, from the above expression for the metric and using the Einstein field equations, one can derive the energy density and pressure of the fluid (see Eqs.~\ref{densityRGexact}, \ref{pressureRGexact}). 

Two  properties of this General Relativity solution should be stressed: i) this system has discontinuous energy density at $r=R$, and continuous but non-differentiable  pressure at the same radius; ii) in order to avoid any  kind of singularity for $r<R$, it is necessary that $M/R < 4/9 = 0.444...$. For $M/R> 1/2$ the ``star'' becomes a black hole, and for  $4/9 < M/R < 1/2$  there is no static solution with finite pressure (see e.g., \cite{0226870332}).

To solve the differential equation (\ref{eq:Lambda1}), a fixing condition for the scale $\mu$ must be used. We assume the ansatz (\ref{eq:mufUUh}) with $G(\mu)$ given by (\ref{eq:Gmu}) and use 
\begin{equation}
    \label{eq:mu1UUh}
    \mu = 1 + U^\alpha U^\beta h_{\alpha \beta} = 1 +U^\alpha U^\beta (g_{\alpha \beta} - \eta_{\alpha \beta}) = - U^0 U^0 \eta_{00} = - \frac 1 {g_{00}} \approx \frac 1 {A(r)},
\end{equation}
where $\eta_{\alpha \beta}$ is a Minkowski metric and we consider that the sphere is at rest with respect to the observer, hence $U^i =0$ was used above. The metric $g_{\alpha \beta}$ is the full spacetime metric, including RG corrections, that is $g_{00} = - A(r) + O(\nu)$. Since $\mu$ only appears multiplied by $\nu$, the field equations up to first order on $\nu$ will not depend on this ``backreaction'' on the $\mu$ setting, hence unless solutions up to second order on $\nu$ are envisaged, the approximation above is sufficient. We stress that the main purpose of the particular setting (\ref{eq:mu1UUh}) is to illustrate the derivation of the $\Lambda$ solution with a nontrivial General Relativity background and the ansatz (\ref{eq:mufUUh}). 

All the necessary information to solve Eq.~(\ref{eq:Lambda1}) was gathered above, and its solution, with the boundary condition $\Lambda(R)=0$ reads
\begin{eqnarray}
\Lambda(r) &= & \frac{24 M \nu}{R^3}  \left[ \ln \frac{2 -4 M /R}{ 4 + ( r^2/R^2 -9)M/R } + 2 \mbox{ arctanh}\left(\frac{\sqrt{1-2 M r^2/R^3}}{3  \sqrt{1 - 2 M/R}}\right)\right]+  \nonumber \\[.2in]
&& + \frac{36  M \nu}{R^3} \, \frac{  1+ \left(r^2/R^2-3 \right)M/R - \sqrt{(1-2 M/R) \left(1-2 M r^2/R^3\right)}}{ 4 + \left(r^2/R^2-9 \right)M/R} + O(\nu^2).
    \label{eq:LsolGRexact}
\end{eqnarray}
By using that, for $x<1$, $\mbox{arctanh } x = \frac 12 \ln \frac{1+x}{1-x}$, it is possible to eliminate the arctanh from the above, but it leads to a larger expression.

The boundary condition used above guarantees that $\Lambda(r)$ is continuous at $r=R$, since outside the sphere $\Lambda$ must be zero according to Eq.~(\ref{eq:lambdaT0null}). 

\begin{figure}[t]
    \centering
    \begin{subfigure}[b]{0.45\textwidth}
            \includegraphics[width= \textwidth]{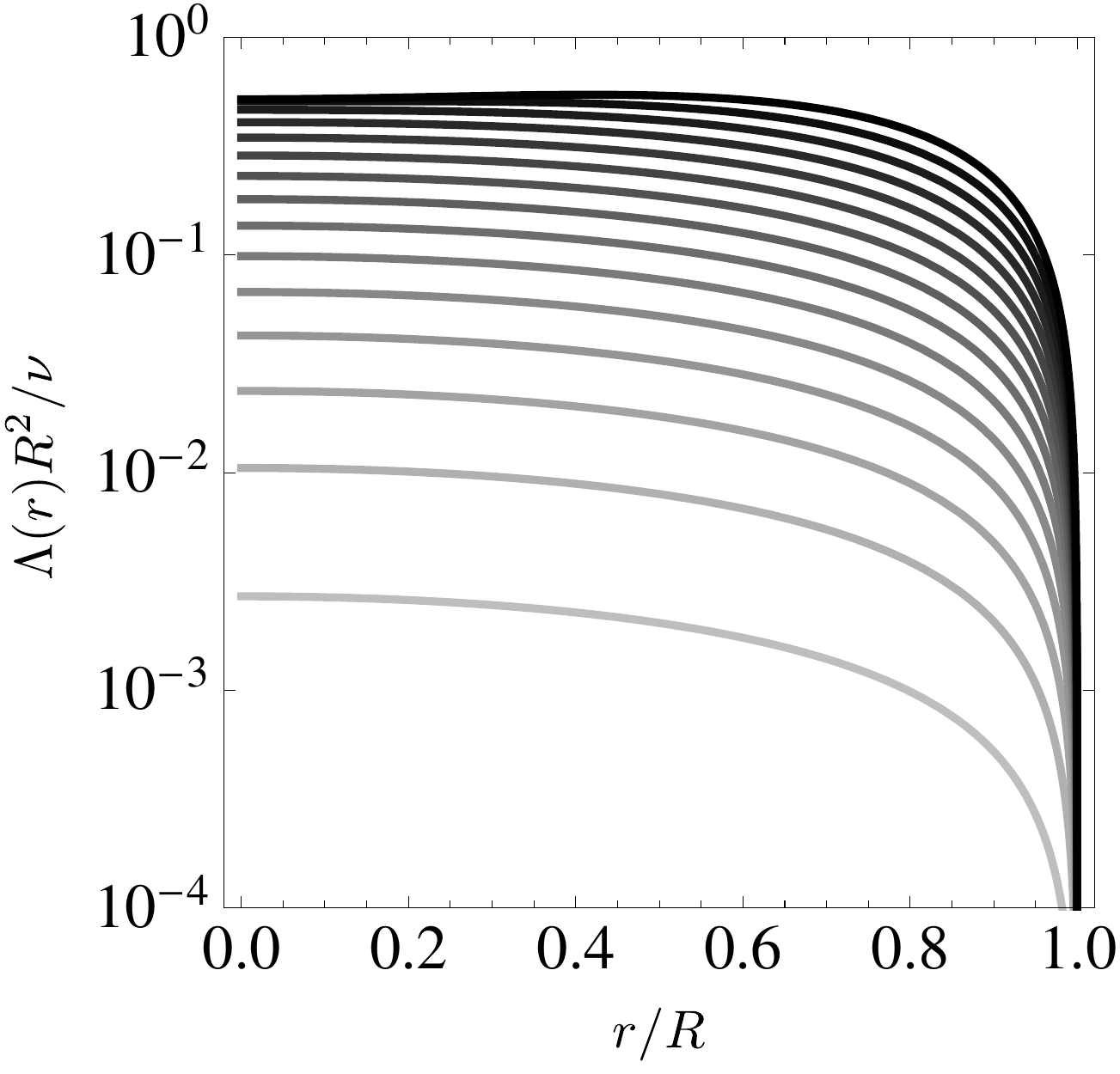}
    \end{subfigure}
    \quad
    \begin{subfigure}[b]{0.45\textwidth}
           \includegraphics[width=0.95\textwidth]{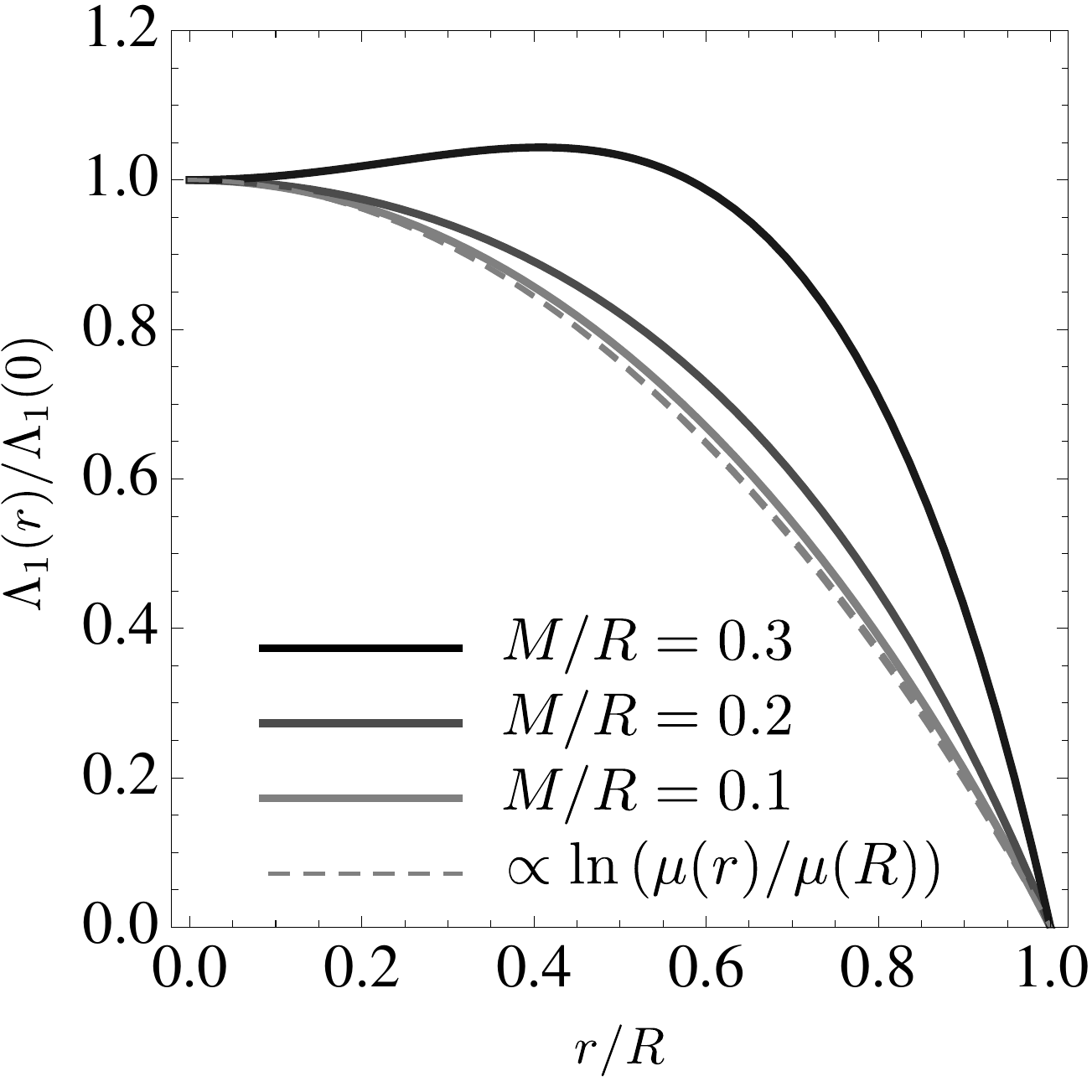}
    \end{subfigure}
  \caption{The behaviour of $\Lambda(r)$ inside a spherical body of constant density and radius $R$, where $\Lambda = \Lambda_0 + \nu \Lambda_1$ and $\Lambda_0=0$. The left plot shows the values of $\Lambda R^2/\nu$ for diverse values of $M/R$, from $M/R=10^{-3}$ (the lightest grey curve) to $M/R=0.3$ (the darkest curve), with a step of $0.02$.  The right plot compares $\Lambda_1(r)/\Lambda_1(0)$, for some values of $M/R$, with a logarithm function of $\mu(r)$ (which was the solution derived for vacuum with G given by Eq. \ref{eq:Gmu}).}
  \label{fig:PlotLinteriorSch}
\end{figure}

Figure \ref{fig:PlotLinteriorSch} shows a plot of the $\Lambda$ solution (\ref{eq:LsolGRexact}) for certain values of $M/R$ and compares it to $\ln \mu$, which was the behaviour derived for $\Lambda_1$ in vacuum when $\Lambda_0 \not=0$. It is shown that as $M/R$ decreases the solution approaches the logarithmic behaviour. This presents a concrete case in which the relation between $\Lambda$ and $\mu$ (or between $\Lambda$ and $G$) changes depending on the matter content of the system. Figure \ref{fig:PlotLinteriorSch} also shows that the induced cosmological constant by the RG approach is roughly constant inside the sphere, and the order of magnitude of $\Lambda R^2/\nu$ inside the sphere is about the same order of $M/R$.\footnote{Assuming $M/R$ smaller and not very close to  $4/9 =0.444...$.} 

\subsection{General Relativity perturbations with Renormalization Group perturbations} \label{RGpertGRpert}

The perturbative picture just presented use exact General Relativity solutions, but such solutions are not always feasible to derive for many applications of physical interest. Hence, in practice, the use of two perturbations may be useful, namely a General Relativistic perturbation in regard to a certain background and a RG perturbation. Therefore, Eqs.~(\ref{eq:perturbRGg}, \ref{eq:perturbRGT}) written explicitly up to first order on both perturbations become
\begin{eqnarray}
    g_{\alpha \beta} &=& \stackrel{(0,0)}g_{\!\!\! \alpha \beta} + \!\!\stackrel{(1,0)}g_{\!\!\!\alpha \beta} +  \nu \!\!\stackrel{(0,1)}g_{\!\!\!\alpha \beta} + ..., \label{eq:perturbRGg2}\\[.1in]
    T_{\alpha \beta} &=& \stackrel{(0,0)}T_{\!\!\! \alpha \beta} + \!\!\stackrel{(1,0)}{T}_{\!\!\!\alpha \beta} + \nu \!\!\stackrel{(0,1)}T_{\!\!\!\alpha \beta} +... \label{eq:perturbRGT2}
\end{eqnarray}
This perturbative scheme is particularly useful if the RG perturbations are considered to be of similar magnitude of the General Relativity ones  (e.g., \cite{Rodrigues:2009vf}). In the above, the first number inside the brackets designates General Relativity perturbation order, and the second number inside the brackets corresponds to the RG one.

In the perturbative scheme above, some care is necessary on the role of $\Lambda_0$. In particular, if one sets the background $ \stackrel{(0,0)}g_{\!\!\! \alpha \beta}$ to be a Minkowski metric, then $\Lambda_0$ is either of the same order of $\nu \Lambda_1$ or smaller. Similarly, in pure General Relativity, if the background is Minkowski the influence of the cosmological constant appears at most at the first perturbative order, likewise the influence of any matter density. In conclusion, within this approach and with $ \stackrel{(0,0)}g_{\!\!\! \alpha \beta}$ as a Minkowski metric, Eq.~(\ref{eq:Lambda1}) can be trivially solved up to first order on the perturbations to yield, with the boundary condition $\Lambda(\mu=1)=\Lambda_0$, $\Lambda_1=0$ and hence
\begin{equation}
    \Lambda = \Lambda_0 + O(\nu^2). \label{eq:lambdaminkowski}
\end{equation}

Before concluding, we remark that if the background $ \stackrel{(0,0)}g_{\!\!\! \alpha \beta}$ is de Sitter, then $ \stackrel{(0,0)}T_{\!\!\! \alpha \beta}=0$ and from  Eq.~(\ref{eq:Lambda1}) one derives Eq. (\ref{eq:lambdaT0null}). In this case, $\Lambda_0$ needs not to be small in any sense.

\subsection{Conformal transformations}\label{sec:conformal}

The existence of certain approximate relation between the RG corrections to the Einstein-Hilbert action and conformal transformations is not a novelty \cite{Reuter:2003ca, Shapiro:2004ch, Rodrigues:2009vf}. Here we review this topic in connection with this paper approaches and results, with particular emphasis on the role of this approximate relation in the presence of matter and cosmological constant. This topic can be approached at the action level, but there are some subtleties that we think can be more easily conveyed at the field equations level, in particular for the non-vacuum case.

By doing the  conformal transformation
\begin{equation}
     \tilde g_{\alpha \beta} = \Omega^2 g_{\alpha \beta},
\end{equation}
the Ricci scalar and the Ricci tensor become respectively \cite{0226870332} 
\begin{eqnarray}
    && \tilde R = \Omega^{-2}(R - 6 \Box \ln \Omega - 6 \nabla_\gamma \ln \Omega \nabla^\gamma \ln \Omega), \\[.1in]
    && \tilde R_{\alpha \beta} = R_{\alpha \beta} - 2 \nabla_\alpha \nabla_\beta \ln \Omega - g_{\alpha \beta} \Box \ln \Omega + 2 \nabla_\alpha \ln \Omega \nabla_\beta \ln \Omega - 2 g_{\alpha \beta}\nabla_\gamma \ln \Omega \nabla^\gamma \ln \Omega. 
\end{eqnarray}
Therefore, by setting $\Omega^2 = G_0/G$, one can map the Einstein tensor $G_{\alpha \beta}$ into the tensor ${\cal G}_{\alpha \beta}$, up to first order on $\nu$, namely
\begin{eqnarray}
    \tilde G_{\alpha \beta}  && \approx G_{\alpha \beta} +  g_{\alpha \beta} \Box \ln (G_0 G^{-1}) - \nabla_\alpha \nabla_\beta \ln (G_0 G^{-1}) \nonumber \\
    && \approx G_{\alpha \beta} +  G_0 g_{\alpha \beta} \Box  G^{-1} - G_0 \nabla_\alpha \nabla_\beta G^{-1} \nonumber \\
    && \approx {\cal G}_{\alpha \beta},
\end{eqnarray}
where second order terms on $\nu$ were neglected. Consequently, if $\tilde g_{\alpha \beta}$ is a solution of the Einstein equations in vacuum, then $g_{\alpha \beta} = \frac{G}{G_0} \tilde g_{\alpha \beta}$ is a solution for 
\begin{equation}
     {\cal G}_{\alpha \beta} \approx 0, \label{eq:Gcalvac}
\end{equation}
up to first order on $\nu$. For vacuum and up to first order on $\nu$, the field equations (\ref{eq:Gcalvac}) are indeed a particular case of the RG extension of General Relativity, in case  $\Lambda$ can be neglected at the same order on $\nu$. The latter case can be realised, for instance, if the integration constant $\Lambda_0$ in the solution expressed in Eq.~(\ref{eq:lambdaT0null}) is zero.

In general, the presence of $\Lambda$ can spoil the conformal mapping between General Relativity and its RG extension, but as remarked in Ref. \cite{Reuter:2003ca}, there is a particular relation between $\Lambda$ and $G$ that can preserve the conformal mapping. It is curious, and we detail this here, that this relation is exactly the one derived for the vacuum case in Eq.~(\ref{eq:lambdaT0null}). Indeed, consider that $\tilde g_{\alpha \beta}$ satisfies the equation
\begin{equation}
    \tilde G_{\alpha \beta} + \tilde g_{\alpha \beta} \Lambda_0 =0.
\end{equation}
Therefore $ g_{\alpha \beta} = \frac{G}{G_0}\tilde g_{\alpha \beta}$, up to first order on $\nu$, is a solution of
\begin{equation}
    {\cal G}_{\alpha \beta} + g_{\alpha \beta} \Lambda \approx 0,
\end{equation}
since $g_{\alpha \beta} G_0 G^{-1} \Lambda_0 \approx g_{\alpha \beta} \Lambda$, from Eq.~(\ref{eq:lambdaT0null}). 

The conformal mapping is also preserved within the perturbative picture about Minkowski spacetime. In this case, $\Lambda_0$ does not contribute to the background metric, hence $g_{\alpha \beta} G_0 G^{-1} \Lambda_0 \approx g_{\alpha \beta} \Lambda_0$. Since $\Lambda=\Lambda_0 + O(\nu^2)$, according to Eq.~(\ref{eq:lambdaminkowski}), the conformal mapping is preserved in the perturbative picture about Minkowski.

\bigskip

 As a last remark on the use of conformal transformations, we consider the presence of a perfect fluid. This issue was briefly studied in Ref. \cite{Rodrigues:2009vf} for a particular case and a different approach. For a review on certain general properties of conformal transformations on fluids, see e.g. \cite{Capozziello:2010zz}. Let
\begin{equation}
    \tilde T_\beta^\alpha =  \tilde \rho \tilde U^\alpha \tilde U_\beta +  \tilde  p \left( \delta^\alpha_\beta  + \tilde U^\alpha \tilde U_\beta\right).
\end{equation}
Using the proper time $\tilde \tau$ as the affine parameter,
\begin{equation}
    \tilde U^\alpha = \frac{dx^\alpha}{d\tilde \tau} = \sqrt{\frac{G}{G_0}} \frac{dx^\alpha}{d \tau} = \sqrt{\frac{G}{G_0}} U^\alpha,
\end{equation}
since $d\tilde \tau^2 = - \tilde g_{\alpha \beta}dx^\alpha dx^\beta  = \frac{G_0}{G}d\tau^2$. Hence, $\tilde U^\alpha  \tilde U_\beta = U^\alpha  U_\beta$.

Let
\begin{equation}
        \label{eq:rhopdef}
       \rho \equiv \left(\frac{G_0}{G}\right)^{n} \tilde \rho \;\;\;\; \mbox{ and } \;\;\;\;  p \equiv \left(\frac{G_0}{G}\right)^{n} \tilde p,
\end{equation}
where $n$ is a constant. Therefore, if $\tilde g_{\alpha \beta}$ is a metric solution for the field equations given by
\begin{equation}
    \tilde G_\alpha^\beta = {8 \pi G_0} \tilde T_\alpha^ \beta,
\end{equation}
then, up to first order on $\nu$, $g_{\alpha \beta} = \frac{G}{G_0} \tilde g_{\alpha \beta}$ is a solution for 
\begin{equation}
    {\cal G}_\alpha^\beta \approx  {8\pi G_0}  \left(\frac{G}{G_0}\right)^{n-1} \left[\rho  U^\alpha  U_\beta +    p \left( \delta^\alpha_\beta  +  U^\alpha  U_\beta\right) \right] = {8\pi G_0} \left(\frac{G}{G_0}\right)^{n-1}  T_\alpha^\beta,
\end{equation}
where it was used that $\tilde G_\alpha^\beta \approx \frac{G}{G_0} {\cal G}_\alpha^\beta$. Clearly, the case of interest corresponds to $n=2$. 

Since the conformal transformation employed here is not a physical process, but just a method to generate solutions, one should not worry on the physical interpretation of the definitions in Eq.~(\ref{eq:rhopdef}). However, it is important to consider whether $\tilde \nabla_\alpha \tilde T^\alpha_\beta=0$ is compatible with $ \nabla_\alpha  T^\alpha_\beta=0$. In general, it is not possible to satisfy both of them, since
\begin{equation}
    \tilde \nabla_\alpha \tilde T^\alpha_\beta =  \left( \frac{G}{G_0}\right)^n \left[ \nabla_\alpha  T^\alpha_\beta +n T^\alpha_\beta \partial_\alpha \ln \left( \frac{G}{G_0}\right) - (\rho + p) \left(\partial_\beta \ln \frac{G}{G_0} + 4 U^\alpha U_\beta \partial_\alpha\ln \frac{G}{G_0}\right) \right]. 
\end{equation}
However, within the perturbative scheme of Eqs.~(\ref{eq:perturbRGg2}, \ref{eq:perturbRGT2}), and if $\stackrel{(0,0)}T_{\!\!\! \alpha \beta}=0$, then indeed $  \tilde \nabla_\alpha \tilde T^\alpha_\beta \approx  \nabla_\alpha  T^\alpha_\beta$ up to first order on the perturbations.

\section{Action and perturbations from a Scalar-Tensor perspective}\label{sec:STcomparison}

\subsection{Actions, field equations and a modified Brans-Dicke approach}\label{sec:STcomparisonOntheactions}

In Eqs. (\ref{eq:RGNOexterConst}, \ref{eq:RGNOexterConstNOmu}) we presented  effective actions for gravity with RG effects such that all their dynamical relevant equations can be straightforwardly derived from the action. Moreover  the notation $\Lambda\{\mu\}$ was introduced in order to stress that $\Lambda$ is not a known function of $\mu$ at the action level. The action (\ref{eq:RGNOexterConstNOmu}) will be useful in particular for a comparison with Brans-Dicke theory.

\subsubsection{Brans-Dicke without kinetic term}

Consider the following Brans-Dicke action with a potential $V$, without a kinetic term for the scalar field (i.e., corresponding to the $\omega=0$ case), and with an action for matter,
\begin{equation}
    S_{\mbox{\tiny BDwM}}[\phi,g, \Psi] = \frac{1}{16 \pi} \int   \phi \left ( R - 2 V(\phi)   \right ) \sqrt{-g} \, d^4x + S_{\mbox{\tiny M}}[g,\Psi].\label{eq:actionBD}
\end{equation}

The field equations read,
\begin{eqnarray}
    && G_{\alpha \beta} + \phi^{-1} \Box \phi g_{\alpha \beta} - \phi^{-1} \nabla_\alpha \nabla_\beta \phi + g_{\alpha \beta} { V(\phi)}= \frac{8 \pi}\phi T_{\alpha \beta}, \label{eq:BDfieldeq1}\\
    &&  R - 2 {V(\phi)} - 2 \phi V'(\phi) = 0, \label{eq:BDfieldeq2} \\
    && \frac{\delta S_{\mbox{\tiny M}}}{\delta \Psi} =0,
\end{eqnarray}
There are different strategies  to look for  the above equations solutions. A straightforward possibility is the following: from the knowledge of a specific potential $V(\phi)$, a energy-momentum tensor $T_{\alpha \beta}$ and certain boundary conditions, one may derive the solutions for $g_{\alpha \beta}$ and $\phi$. In the end this procedure determines these fields as spacetime functions. Depending on the nature of the problem, one can try to invert the problem by fixing $V(\phi)$ from the knowledge of some expected behaviour  of the metric. Nonetheless, once $V(\phi)$ is determined, it should be universal, that is, if $V(\phi) = \phi^2$ for cosmology, it should also be $V(\phi) = \phi^2$ for the solar system. On the other hand, we remark that any relation between the Brans-Dicke field $\phi$ and the metric (or other fields) is circumstancial, that is, is system dependent. For instance, for a given potential $V(\phi)$, a given $T_{\alpha \beta}$ and given boundary conditions for $g_{\alpha \beta}$ and $\phi$, there may exist some algebraic relation between $\phi$ and other fields, but there is no reason for  such relation to be preserved if either $T_{\alpha \beta}$ or the boundary conditions are changed, since there is no ``constraint'' that imposes such possible relation.

\subsubsection{Brans-Dicke with a constrained field and system-dependent potential}

If one uses the Eqs.~(\ref{eq:BDfieldeq1}, \ref{eq:BDfieldeq2}) but considers that the potential $V(\phi)$ is an unknown function of $\phi$, it is not possible in general to determine all the three unknowns, namely $\phi$, $g_{\alpha \beta}$ and provide a specific dependence of $V$ on $\phi$. In this picture, one has all the unknowns of a Brans-Dicke theory plus one.

Consider that the potential is an unknown function of $\phi$ at the action level (i.e., $V(\phi) \rightarrow V\{\phi\}$), and a constraint is imposed, such that it asserts a fixed  relation between $\phi$ and other fields, namely, 
\begin{equation}
    S_{\mbox{\tiny ModBDwM}}[\phi,g, \lambda,\Psi] = \frac{1}{16 \pi} \int   \left [ \phi  \left ( R - 2  V\{\phi\}  \right ) + \lambda (\phi - F(g,\Psi)) \right] \sqrt{-g} \, d^4x + S_{\mbox{\tiny M}}[g,\Psi]. \label{eq:actionModBDwL}
\end{equation}
The modified Brans-Dicke action above is identical to the RG gravity extension expressed in Eq.~(\ref{eq:RGNOexterConstNOmu}), apart from the replacements $\phi \rightarrow G^{-1}$ and $V\{\phi\} \rightarrow \Lambda\{G\}$. For simplicity, assuming that $F$ does not depend on derivatives on $g_{\alpha \beta}$ or $\Psi$, the field equations read\footnote{In case F depends on derivatives on the metric or on the $\Psi$ fields, then certain straightforward replacements are necessary, for instance $\lambda \frac{\partial F}{\partial g^{\alpha \beta}} \rightarrow \frac{1}{\sqrt{-g(x)}}\int \lambda(y) \frac{\delta F(y)}{\delta g^{\alpha \beta}(x)} \sqrt{-g(y)} d^4y$.}
\begin{eqnarray}
    && G_{\alpha \beta} + \phi^{-1} \Box \phi \, g_{\alpha \beta} + \phi^{-1} \nabla_\alpha \nabla_\beta \phi + g_{\alpha \beta} V  = \frac{8 \pi}\phi T_{\alpha \beta}  +\lambda \frac{\partial F}{\partial g^{\alpha \beta}}, \label{eq:ModBD1}\\
    && R - 2 (V \phi)' + \lambda =0, \label{eq:ModBD2}\\
    && \phi = F(g,\Psi), \label{eq:ModBD3}\\
    && \lambda \frac{\partial F}{\partial \Psi} = \frac{1}{\sqrt{-g}}\frac{\delta S_{\mbox{\tiny M}}}{\delta \Psi}, \label{eq:ModBD4}
\end{eqnarray}
where the prime denotes a derivative with respect to $\phi$. The above equations need not to satisfy $\nabla^\alpha T_{\alpha \beta}=0$. This can be straightforwardly verified from the field equations above, and it is also expected from the action (\ref{eq:actionModBDwL}), due to the term $\lambda(\phi - F)$, which leads in general to $\delta S_M [g,\Psi]/ \delta \Psi \not= 0$.

Since $\Psi$ is a set of fields of any tensorial nature, if there is at least a field, say $\psi$, such that the right hand side of Eq.~(\ref{eq:ModBD4}) is zero and $\partial F/\partial \psi \not=0$, then $\lambda =0$. A particular realisation of such $\psi$ field comes from the ansatz (\ref{eq:mufUUh}), where the field $\gamma_{\alpha \beta}$ plays the role of this $\psi$ field. In case the condition $\lambda=0$ is met at the field equations level, the field equations become
\begin{eqnarray}
    && G_{\alpha \beta} + \phi^{-1} \Box \phi \, g_{\alpha \beta} - \phi^{-1} \nabla_\alpha \nabla_\beta \phi + g_{\alpha \beta} V   = \frac{8 \pi}\phi T_{\alpha \beta}, \label{eq:ModBD1a}\\
    && R - 2 (V \phi)' =0, \label{eq:ModBD2a}\\
    && \phi = F(g,\Psi),\label{eq:ModBD3a}\\
    && \frac{\delta S_{\mbox{\tiny M}}}{\delta \Psi}=0. \label{eq:ModBD4a}
\end{eqnarray}
From the first two equations above it is possible to infer that $\nabla^\alpha T_{\alpha \beta}=0$ (the proof follows the same steps presented for Eq.~(\ref{eq:consist}) derivation).  It should be clear that the above equations have the same form of the field equations derived from the RG action (\ref{eq:RGNOexter}), together with a scale setting condition for $\mu$. Equations (\ref{eq:ModBD1a}, \ref{eq:ModBD2a}) come directly from that action (apart from trivial change of names, and assuming that $\mu$ can be written as a function of $G$),  Eq.~(\ref{eq:ModBD3a}) corresponds to the scale setting condition, and Eq.~(\ref{eq:ModBD4a}) is just a statement that the matter field equations are not directly affected by the RG effects in gravity.

The form of the field equations (\ref{eq:ModBD1}-\ref{eq:ModBD4a}) do not depend on whether one uses $V(\phi)$ or $V\{\phi\}$, but the meaning of these differential equations, and their solutions, depend on these choices. Within the usual approach, there is a potential that is a known function of $\phi$ at the action level, $V(\phi)$. Hence, in Eq.~(\ref{eq:ModBD2a}),  $(V \phi)'$ is a  just a known function of $\phi$; that is, this equation is stating that $R$ can be seen as a function of $\phi$.  If there is an inverse, one can express $\phi$ as a function of $R$, and the action (\ref{eq:actionModBDwL}), apart from the constraint term which will be discussed latter, becomes the well known $f(R)$ action. The relation between Brans-Dicke with no kinetic term and a potential (also Minimal Dilatonic Gravity \cite{Fiziev:2012js})  and $f(R)$ theories is known for a long time, for reviews on this demonstration see, e.g., \cite{Sotiriou:2008rp, Rodrigues:2011zi}. 

The above relation with $f(R)$ was achieved by both assuming the existence of an inverse for $(V \phi)'$ and ignoring the constrain, that is without using Eq.~(\ref{eq:ModBD3a}), which is an independent equation. From the RG perspective, this relation comes from QFT arguments that need not to have any relation with Eq.~(\ref{eq:ModBD2a}). Hence, if $(V \phi)'$ is a given $\phi$ function that can be inverted,  Eq.~(\ref{eq:ModBD3a}) is in general incompatible with  Eq.~(\ref{eq:ModBD2a}). 

For the case $V = V\{\phi\}$, there is no reason for Eqs.~(\ref{eq:ModBD2a}, \ref{eq:ModBD3a}) be incompatible. In particular, one may interpret Eq.~(\ref{eq:ModBD2a}) as a differential equation whose solution will fix the relation between $V$ and $\phi$ for a given system (this procedure is exemplified in section~\ref{sec:Lsol}). It is correct to state that, for a given system, $R$ is some function of $\phi$ due to Eq.~(\ref{eq:ModBD2a}), but contrary to the previous case such function is unknown beforehand, system dependent and for many cases it is noninvertible. Actually there is no need for $V$ to be an analytical function of $\phi$ for a given system (which is a usual requirement for a potential $V(\phi)$). Some aspects of the two latter points, the system dependence and inversibility, can be found in section \ref{sec:Lsol}, but below these points are developed in a more specific way.
 
 The system dependent features are easily seen if Eq. (\ref{eq:ModBD2a}) is re-written in a form analogous to (\ref{eq:consistrho}), that is by eliminating $R$. From the trace of (\ref{eq:ModBD1a}),
\begin{equation}
    R =  - \frac{8 \pi}{\phi} T + 3 \phi^{-1} \Box \phi + 4 V,
\end{equation}
hence Eq. (\ref{eq:ModBD2a}) is equivalent to
\begin{equation}
     \phi^2 V' -  \phi V = - 4 \pi  T + \frac 32  \Box \phi. \label{eq:difV1}
\end{equation}
The relation between $V$ and $\phi$ in the standard scalar tensor picture is an off-shell relation, that is, a relation that is valid in spite of the field equations. The above equation, can be interpreted as providing an on-shell relation between $\phi$ and the other fields. In this usual picture, the left-hand side is a known $\phi$ function, while the right hand side cannot {\it a priori} be put in a form of a $\phi$ function, an hence this equation provides a on-shell relation between $\phi$, $\Box \phi$ and $T$. In the proposed picture where $V$ is $V\{\phi\}$, the left hand side is an unknown function of $\phi$, while the right hand side satisfies a relation with a function of $\phi$. This relation is in part provided by Eq. (\ref{eq:ModBD3a}), which is absent in the standard Brans-Dicke case. Hence in the $V\{\phi\}$ picture, which is formally equivalent to the RG approach described previously, the relation between $V$ and $\phi$ is an on-shell relation, and as such it is subject to changes due to changes in the matter content. In a certain region in the universe $V$ could be closer to $\phi^2$, and in another region, say, $\phi^4$ (or even be two different non-analytical solutions in different regions). Moreover, in the standard picture $V(\phi)$ is a $\phi$ function completely independent on the matter part, while in the system-dependent case $V\{\phi\}$ is a function of $\phi$ that can depend on constants that are associated with matter properties, hence two different systems can yield different solutions for $V\{\phi\}$.

 There are some known exact solutions for the two first equations, which constitute Brans-Dicke exact solutions; but the difficult relies on deriving exact solutions that both satisfy the two first equations together with a nontrivial $F$ function for Eq. (\ref{eq:ModBD3a}).  The next subsection enters in further details on the issue of solving these equation with a perturbative method, and its correspondence with section \ref{sec:Lsol}.

\subsection{Perturbative schemes and their comparison}

In section \ref{sec:Lsol} it was shown how to derive perturbative solutions for $\Lambda\{G\}$. The steps are exactly the same for the derivation of $V\{\phi\}$, since the RG approach with unknown $\Lambda\{G\}$ and a fixed relation between $G$ and other fields is formally the same problem of the proposed modified Brans-Dicke model. The purpose of this subsection is to invert the problem, putting it in a standard ST picture, where one starts from the knowledge of a potential $V$ and derives the solution for $\phi$. In order to proceed in this way, the potential that will be used at the action level will be the potential $V\{\phi\}$ (or $\Lambda\{G\}$) that was derived for vacuum. Note that this potential is  a function of $\phi$, but it has an unusual property, namely the constants it depends on are related with matter properties. This procedure is done both to check the correspondence with standard Brans-Dicke and to clarify the map between the perturbations of the different theories.

\subsubsection{Vacuum}

The starting point is the Brans-Dicke action (\ref{eq:actionBD}) with $S_M =0$ and it is assumed that $\phi$ does not change much in regard to a certain background value which is a constant named $\phi_0$. That is, by defining 
\begin{equation}
     \varphi \equiv \frac{\phi}{\phi_0} - 1,
\end{equation}
the assumption is that  $|\varphi| \ll 1$. Due to the above definition, the equality $\phi = \phi_0(1  + \varphi)$ is an exact one, that is, it is valid beyond the first order perturbation on $\varphi$.

The potential is chosen to be 
\begin{equation}
    V(\phi) = \Lambda_0 \frac{\phi}{\phi_0} +  O(\varphi^2), \label{eq:potential1st}
\end{equation}
which corresponds to the $\Lambda$ derived for vacuum in Eq.~(\ref{eq:lambdaT0null}), apart from the notation change.

The field equations that the metric and $\phi$ must satisfy are given by Eqs.~(\ref{eq:BDfieldeq1}, \ref{eq:BDfieldeq2}) with $T_{\alpha \beta}=0$. Equation (\ref{eq:BDfieldeq1}) up to the zeroth order on $\varphi$ is simply General Relativity in vacuum and with cosmological constant $V_0$. Equation (\ref{eq:BDfieldeq2}) can be re-written as Eq.~(\ref{eq:difV1}), which in turn it is also equivalent to (assuming $\nabla_\alpha \phi \not=0$),\footnote{In the RG framework, Eq. (\ref{eq:difV2}) corresponds to Eq.~(\ref{eq:consistrho}).}
\begin{equation}
      \nabla_\alpha V  = \left (- \frac{4 \pi}{\phi} T + \frac 32 \phi^{-1} \Box \phi +  V \right) \phi^{-1}\nabla_\alpha \phi.\label{eq:difV2}
\end{equation}
Equations (\ref{eq:difV1}, \ref{eq:difV2}) in exact form are redundant with energy-momentum tensor conservation, which is always true in vacuum ($T_{\alpha \beta} =0$). From a perturbative perspective, it should be clear that Eq.~(\ref{eq:difV1}) is  nontrivial  even at zeroth order on $\varphi$. The reason being that it depends on the derivative of a function with respect to $\phi$. Naturally, if one multiply both sides of that equation by $\nabla_\alpha \phi$, the term $V'$ can be replaced by $\nabla_\alpha V$ , see Eq.~(\ref{eq:difV2}), and the equation becomes trivial at zeroth order on $\varphi$. In the end both approaches are consistent and yield the same result. 

For $T_{\alpha\beta} =0$, Eq.~(\ref{eq:difV1}) up to zeroth order on $\varphi$, and Eq.~(\ref{eq:difV2}) up to first order on $\varphi$, are both satisfied for any $\phi$ if the potential $V$ is given by (\ref{eq:potential1st}). It is  curious that the potential (\ref{eq:potential1st}) has the property of leading to solutions that are valid for arbitrary $\phi$ (up to first order on $\varphi$). The reason for this behaviour associated with this particular potential can be  traced to the conformal transformation relation presented in section~\ref{sec:conformal}.

\subsubsection{Sphere of constant density}

As a second example, we consider the solution inside a uniform density matter considered in section~\ref{sec:RGpertGRexactInterior}. In order to convert that solution into the notation and procedures of a Brans-Dicke theory, the task is to convert the $\Lambda(r)$ expression (\ref{eq:LsolGRexact}) into a potential $V(\phi)$. This procedure was trivial in the previous example since we already knew how to express $\Lambda$ and a function of $G^{-1}$. We divide this task in the following steps: i) since $\mu = 1/A$ (according to Eq.~(\ref{eq:mu1UUh})), we will first express $r$ as a function of $A$; ii) express $r$ as a function of $G^{-1}$; iii) from the knowledge of $\Lambda(r)$, we can therefore express $V(\phi)$ (which is $\Lambda(G^{-1})$, apart from a notation change). 

Since the Brans-Dicke picture is also perturbative, and in some sense up to first order, it is tempting to assume that the final expression for $V(\phi)$ should be a linear function on $\phi$, but this is not necessary, and this example will show this. The essential point is that although $\phi/\phi_0 - 1$ is a first order term on $\varphi$, the term $(\phi/\phi_0 - 1)/\nu$, where $\nu$ is a small constant, can be much greater than the former if $\nu$ is sufficiently small.

Equation (\ref{eq:A(r)}) defines the function $A(r)$. To simplify the notation, we choose to measure both the distance $r$ and the ``mass'' $M$ in $R$ unities, which effectively is the same of replacing  $R$ by $1$ in the equation that defines $A$, and to be even more economic we simply set $R=1$ (which implies that both $M$ and $r$ are now dimensionless). It should be clear that the function $A$ does not have an inverse for arbitrary $M$ values. Actually, considering the previous discussion on $V\{\phi\}$ and $\Lambda\{G^{-1}\}$, one should not expect to be able to write $V$ as a function of $\phi$ for arbitrary values of the parameters that describe matter. In particular, for $r \le 1$ and $0 \le M < 4/9 = 0.444...$, it is possible to write $r$ as a function of $A$, namely
\begin{equation}
    r = \sqrt{\frac{1 - 4 \left ( \frac 32 \sqrt{ 1 - 2 M} - \sqrt A\right)^2}{2M}}.
\end{equation}
Indeed, by solving the equation $A'(r) =0$ for $M$, one derives that either $M=0$ or $M = 4/(9-r^2)$. Since $0<r<1$,  $A(r)$ is a monotonic function for $M<4/9$ (otherwise $A$ will have a local minimum).

The next step is to express $r$ as a function of $G_0 G^{-1} = 1 - 2 \nu \ln A$, since $A = 1/\mu$. Hence, it is only necessary to replace $A$, in the above equation, by $\exp[(1 - G_0 G^{-1})/2\nu]$, or equivalently, by\footnote{It should be stressed that $\ln A$ needs not to be small in some sense, it is $\nu \ln A$ which needs to be small, and this was used in the RG approach in section \ref{sec:Lsol}.} $\exp[(1 - \phi / \phi_0)/2\nu] = \exp (- \varphi/2\nu)$. And the potential reads
\begin{eqnarray}
    V(\phi) & = &  24 M \nu \left [ \ln \frac{1 - 2 M}{3 \sqrt{A (1 - 2 M)} -  A}  + 2 \, \mbox{arctanh} \left (1 - \frac{ 2\sqrt A   }{ 3 \sqrt{1-2M}} \right) \right ] + \nonumber\\[0.2in]
    && + 12 M \nu \frac{  6 \sqrt{A(1-2M)} + 6M - \sqrt{1-2M} ( 3 \sqrt{1-2M} - 2 A) - 3 - 2 A}{  2 \sqrt{A (1-2M)}-\frac 23 A} + O(\nu^2), \nonumber \\[.3in]
    &=& 12 M \nu  \left ( 3 +  2 \ln(1-2M) + \tilde{\varphi}  - 3 \, e^{\frac{\tilde \varphi}{4}} \sqrt{1-2M}  \right ) + O(\nu^2) \label{eq:Vphiexact} 
\end{eqnarray}
In the above it was used that, for $x<1$, $\mbox{arctanh } x = \frac 12 \ln \frac{1+x}{1-x}$,  and $A$, $\varphi$ and $\tilde \varphi$ should be seen as the following shorthand notations: $A = e^{{-\varphi}/({2\nu})}$, $\varphi = \phi/\phi_0 -1$ and $\tilde \varphi = \varphi/\nu$.

In order to clarify the correspondence between the $\Lambda(r)$ expression given in Eq.~(\ref{eq:LsolGRexact}) and the $V(\phi)$ expression presented above, one can do a map between certain general properties that $\Lambda(r)$ has and $V(\phi)$ must also posses. Firstly, $\Lambda(r)$ is zero at zeroth order on $\nu$ and it is non null at first order. Indeed the same property can be spotted in $V(\phi)$ if $\varphi$ is a perturbation of the $\nu$ order. Indeed, if one writes the $\phi$ perturbation ($\varphi$) as  $ \varphi = \nu \tilde \varphi$, assuming that $|\nu \tilde \varphi| <1$ and $|\nu| <1 $, then Eq.~(\ref{eq:Vphiexact}) is clearly a first order $\nu$ expansion in which the zeroth order term is zero. It can be easily spotted that both $\Lambda$ and $V$ are null functions if $M$ is zero. From a standard Brans-Dicke perspective, from one side the potential given above is reasonable since the only field it depends is on $\phi$, on the other hand it has the peculiar feature that the constants it depends happen to coincide with the constants that describe the matter distribution. Another important property is that $\Lambda(r)$ satisfies the boundary condition $\Lambda(1)=0$ (using $R=1$). To show that $V(\phi)$ satisfies the same boundary condition but with a different notation, first one should note that  $r=1$ implies $A = 1 - 2M$. Let $\phi_1$ be the value of $\phi$ that corresponds to $r=1$, then  $\phi_1 = \phi_0(1 - 2\nu\ln (1 -2M))$, and  indeed it is easy to check that $V(\phi_1)=0$.

Now that $V(\phi)$ is known for a relevant range on $M$, the next task is to apply it into Eq.~(\ref{eq:difV1}). In conformity with the perturbative picture, within the lowest order on $\nu$, and using the potential (\ref{eq:Vphiexact}), Eq.~(\ref{eq:difV1}) becomes
\begin{equation}
    12  M - 9 e^{\tilde \varphi/4} \sqrt{1-2M}M \approx - \frac{4 \pi}{\phi_0} \stackrel{(0)}T. \label{eq:zerothorderBDV}
\end{equation}
To do the derivation above, in particular one can use $\partial_{\phi} V = \partial_{\tilde \varphi}V  /(\nu \phi_0)$, hence the term $\phi^2 V'$ contributes with zeroth order terms (these appear in the above equation), while $\phi V$ has no zeroth order contribution. On the right hand side of the above equation, only $T = \stackrel{(0)}T + \stackrel{(1)}T +...$ can contribute with zeroth order terms, hence $\Box \phi$ should be neglected within this approximation. Since $\stackrel{(0)}T$ is the trace of the energy-momentum tensor of the General Relativity solution for the interior of a sphere of constant density, $T = -\rho + 3 p$, with \cite{9780521829519, 0226870332}
\begin{eqnarray}
    \rho &=&  \frac{3 M}{4 \pi}, \label{densityRGexact} \\[.2in]
    p &=& 3 \rho \frac{\sqrt{1 - 2M r^2} - \sqrt{1-2M}}{3 \sqrt{1-2M} - \sqrt{1-2M r^2}}. \label{pressureRGexact}
\end{eqnarray}
Combining the above, Eq.~(\ref{eq:zerothorderBDV}) can be solved to yield
\begin{equation}
    \tilde \varphi = - 4 \ln \left( \frac 32 \sqrt{1-2M} - \frac 12 \sqrt{1-2Mr^2} \right),
\end{equation}
and hence
\begin{equation}
    \frac{\phi}{\phi_0} = 1 + 2 \nu \ln \frac{1}{A(r)},
\end{equation}
where $A(r)$ is given by Eq.~(\ref{eq:A(r)}). The above expression for $\phi$ clearly coincides with the expression for $G(\mu)$ and the scale setting used for $\mu$ which lead to the $\Lambda$ expression in Eq. (\ref{eq:LsolGRexact}). In conclusion, in the above it is illustrated in detail how one can start with the  cosmological constant that is derived from the RG approach, interpret it as a potential within standard ST gravity,\footnote{The only unusual issue in this potential is its dependence on constants that coincide with certain matter distribution properties. This potential is a standard ST potential, hence it is fixed at the action level, and if the matter distribution changes for some reason, the form of the potential will not change.} and then derive $G(\mu(x))$, hence inverting the problem. The above also shows how to map the perturbative RG effects into the perturbative ST case.

\section{Conclusions} \label{sec:conclusions}

The running of $G$ and $\Lambda$ at astrophysical scales, within the framework of the Renormalization Group (RG), is being currently considered from a number of different approaches \cite{Reuter:1986wm, Goldman:1992qs, Bertolami:1993mh, Bottino:1995sa, Bonanno:2001hi, Reuter:2003ca, Reuter:2004nv, Shapiro:2004ch, Reuter:2004nx, Babic:2004ev, Bauer:2005rpa, Brownstein:2005zz, Grande:2007wj, Borges:2007bh, Rodrigues:2009vf, Grande:2010vg, Domazet:2010bk, Domazet:2012tw, Koch:2010nn, Cai:2011kd, Ahn:2011qt, Costa:2012xw, Hindmarsh:2012rc, Perico:2013mna, Rodrigues:2014xka, Lima:2014hia, Koch:2014joa}. It extends General Relativity (GR) and such effect needs not to be negligible. In particular, its magnitude may be sufficient to become important to the proper understanding of either dark matter or dark energy.

Notwithstanding the interest on these theories, the differences between the RG extensions to GR and the probably most natural and well known covariant extension of GR, namely standard Scalar-Tensor (ST) gravity, has not received due attention. Among the three large classes of approaches to the implementation of RG effects in gravity (i.e., the improved action, improved field equations or the improved solutions \cite{Reuter:2003ca}), here we studied the case that is most similar to ST gravity, namely the improved action case. The latter  started to be studied and compared with Brans-Dicke-like actions in Ref. \cite{Reuter:2003ca}, and the main difference was said to be due to the use of external scalar fields. For the simplest and probably more natural case in which the scalar field $G$ does not appear with derivatives (see Eqs.~\ref{eq:RGexter} and \ref{eq:RGNOexter}), Refs. \cite{Koch:2010nn, Domazet:2012tw, Hindmarsh:2012rc} have shown that there is no need to use external scalar fields in order to derive the same field equations, and that this approach can be reduced to  a particular class of $f(R)$ gravity (considering also that $\nabla_\alpha T^{\alpha \beta}=0$ and $\nabla_\alpha \mu \not=0$ everywhere, if the latter condition is dropped, see Ref. \cite{Chauvineau:2015cha}). In this picture, the RG scale $\mu$ is not specified, but derived from the field equations and it is always a function of the Ricci scalar $R$. On the other hand, the approach used by some of us in Refs. \cite{Rodrigues:2009vf, Rodrigues:2012qm, Rodrigues:2014xka} is consistent with Ref. \cite{Reuter:2003ca}, but cannot be reduced to an $f(R)$ gravity.

Here it is shown that if one starts from a  motivation for  $\beta_G$ (the $\beta$-function of $G$) and for the RG scale $\mu$ identification, as the one proposed in section \ref{sec:scalesettingproposal}, then the action to be used is given by Eq.~(\ref{eq:RGNOexterConst}), or equivalently by Eq.~(\ref{eq:RGNOexterConstNOscalar}), where the running of $\Lambda$ is not universal or fixed at the action level, but it is inferred from the field equations. In this case, the RG extended GR needs not to be either a class of $f(R)$ theories or even of standard ST theories. This  approach that is developed here is consistent with Ref. \cite{Reuter:2003ca}, and extends it in the sense that all the necessary information is now present in the action, and no external fields are used. In particular, this is an important step in order to address comparisons to ST gravity.

The use of ST theories with system-dependent potentials, in spite of its connection with the RG effects, can be a relevant approach by itself.   General results valid for any ST theory for a given system are automatically valid for the system-dependent potential case. We have shown in section \ref{sec:STcomparison} how to map solutions between the standard and the system-dependent cases within different perturbative schemes. We stress that, while the standard case is defined from the specification of the function $V(\phi)$ at the action level, and the relation between $\phi$ and other fields is circunstancial, for the system dependent case it is the relation between $\phi$ and the other fields that is specified at the action level, and the relation between $V$ and $\phi$  is circunstancial. Moreover, a natural perturbative scheme in one picture does not lead necessarily to a natural perturbative scheme in the other picture. In conclusion, these approaches lead in general to different physical results once different systems are contrasted. A possible application of this ST structure is for screening mechanisms, e.g., \cite{Vainshtein:1972sx, Khoury:2003aq, Hinterbichler:2010es, Koivisto:2012za}, which we plan to address in a future work.

\acknowledgments

The authors thank Ilya Shapiro for reviewing part of this work and for discussions on the Renormalization Group, and J\'ulio Fabris for discussions on the Ray's fluid description. We also thank the program ``Science without Borders'' (CNPq) that has partially supported visits to the University of Esp\'irito Santo (Brazil) and Observatoire de la C\^ote d’Azur (France). DCR and OFP thank  CNPq (Brazil) for partial financial support. DCR also thanks FAPES (Brazil) for partial financial support.

\appendix

\section{Renormalization Group improved Einstein-Hilbert action as a  metric theory of gravity with nonminimal coupling to matter} \label{app:elimination}

For completeness, we show here that the actions (\ref{eq:RGNOexterConst}) and (\ref{eq:RGNOexterConstNOscalar}) ($S$ and $\bar S$) generate the same field equations. We point that the term $\mu - f(g,\Psi)$ is not a constrain in the strict sense, since (although not explicit) $f$ may also depend on an arbitrary number of derivatives on the metric $g_{\alpha \beta}$ and on the matter fields $\Psi$, that is to  fully explicit the $f$ dependence one should write $f(g, \Psi, \partial g, \partial \Psi, \partial \partial g, \partial \partial \Psi, \,  ...)$.

First we write the actions $S$ and $\bar S$ as follows,
\begin{eqnarray}
	S[g,\lambda, \mu,\Psi] &=& S_g[g,\mu] + \int \lambda (\mu - f(g,\Psi))\,  d^4x  + S_m[g,\Psi], \\[.1in]
	\bar S [g, \Psi] &=& \bar S_g [g, \Psi] +  S_m[g,\Psi].
\end{eqnarray}

The field equations of $S$ can be expressed as,\footnote{On the notation used for these equations, see for instance the chapter on classical fields of Ref.~\cite{9783540591795}.}
\begin{eqnarray}
	\frac{\delta S}{\delta g^{\alpha \beta}(x)} &=& \frac{\delta S_g}{\delta g^{\alpha \beta}(x)} + \int \lambda(y) \frac{\delta \left [ (\mu(y) - f(y)) \sqrt{- g(y)}\right]}{\delta g^{\alpha \beta}(x)}d^4y - \frac{\sqrt{-g(x)}}{2} T_{\alpha \beta}(x) =0, \label{eq:S/g}\\[.2in]
	\frac{\delta S}{\delta \mu(x)} &=& \frac{\delta S_g}{\delta \mu(x)} + \sqrt{-g(x)} \lambda(x) =0, \label{eq:S/mu} \\[.2in]
	\frac{\delta S}{\delta \lambda(x)} &=& \mu(x) - f(x) =0, \label{eq:S/lambda}\\[.2in] 
	\frac{\delta S}{\delta \Psi(x)} &=& - \int \lambda(y) \frac{\delta f(y)}{\delta \Psi(x)} \sqrt{-g(y)} d^4y + \frac{\delta S_m}{\delta \Psi(x)}=0. \label{eq:S/Psi}
\end{eqnarray}
By using Eqs.~(\ref{eq:S/mu}, \ref{eq:S/lambda}), the Eqs.~(\ref{eq:S/g}, \ref{eq:S/Psi}) can be written as
\begin{eqnarray}
	\frac{\delta S}{\delta g^{\alpha \beta}(x)} &=& \frac{\delta S_g}{\delta g^{\alpha \beta}(x)} + \int \frac{\delta S_g}{\delta \mu(y)} \frac{\delta  f(y)}{\delta g^{\alpha \beta}(x)} \sqrt{- g(y)}\, d^4y - \frac{\sqrt{-g(x)}}{2} T_{\alpha \beta}(x) =0, \label{eq:S/g2} \\[.2in]
	\frac{\delta S}{\delta \Psi(x)} &=& \int \frac{\delta S_g}{\delta \mu(y)}\frac{\delta  f(y)}{\delta \Psi(x)}\, d^4y + \frac{\delta S_m}{\delta \Psi(x)}=0. \label{eq:S/Psi2}
\end{eqnarray}

On the other hand, the field equations of $\bar S$ (whose dependence can also be expressed as $\bar S[g, f(g,\Psi)]$) read
\begin{eqnarray}
		\frac{\delta \bar S}{\delta g^{\alpha \beta}(x)} &=& \frac{\tilde \delta \bar S_g}{\tilde \delta g^{\alpha \beta}(x)} + \int \frac{\delta \bar S_g}{ \delta f(y)}\frac{\delta f(y)}{ g^{\alpha \beta}(x)}\, d^4y- \frac{\sqrt{-g(x)}}{2} T_{\alpha \beta}(x) =0,\label{eq:Sbar/g}\\[.2in]
		\frac{\delta \bar S}{\delta \Psi(x)} &=& \int \frac{\delta \bar S_g}{\delta f(y)}\frac{\delta  f(y)}{\delta \Psi(x)}\, d^4y + \frac{\delta  S_m}{\delta \Psi(x)}=0, \label{eq:Sbar/Psi}
\end{eqnarray}
where ${\tilde \delta \bar S}/{\tilde \delta g^{\alpha \beta}}$ is a variational derivative that only considers explicit terms on $g^{\alpha \beta}$, hence it does not consider the metric terms inside $f$. In other words, the full variation with respect to the metric is given by the first two terms in Eq.~(\ref{eq:Sbar/g}).

To conclude, we note that Eqs.~(\ref{eq:S/g2}, \ref{eq:S/Psi2}) (together with Eq. \ref{eq:S/lambda}) are equivalent to Eqs.~(\ref{eq:Sbar/g}, \ref{eq:Sbar/Psi}), and hence $S$ and $\bar S$ generate the same field equations and are classically equivalent. It is assumed that all the surface terms coming from integrations by parts are zero.

\bibliographystyle{JHEP} 

\bibliography{bibdavi2014B}{}

\end{document}